\documentclass[12pt]{article}

\usepackage{amsthm,graphicx,cite,lscape,float}
\usepackage{epsf,latexsym}
\usepackage{amssymb,amsmath}

\newcommand{\rmi}{{\rm i}}
\newcommand{\rme}{{\rm e}}
\newcommand{\rmd}{{\rm d}}
\newcommand{\gs}{|z_{\text{\tiny R}}\rangle}

\newcommand{\zr}{z_{\text{\tiny R}}}
\newcommand{\gr}{\Gamma _{\text{\tiny R}}}

\newcommand{\er}{E_{\text{\tiny R}}}

\newcommand{\Cr}{c_{\text{\tiny R}}}
\newcommand{\Co}{c_{\text{\tiny R,1}}}
\newcommand{\Ct}{c_{\text{\tiny R,2}}}
\newcommand{\nr}{N_{\text{\tiny R}}}
\newcommand{\kr}{k_{\text{\tiny R}}}
\newcommand{\ar}{\alpha_{\text{\tiny R}}}
\newcommand{\br}{\beta_{\text{\tiny R}}}

\begin{document}

\title{\bf Numerical calculation of the decay widths, the decay constants, 
and the decay energy spectra of the resonances of the delta-shell potential}

\author{Rafael de la Madrid \\
\small{\it Department of Physics, Lamar University,
Beaumont, TX 77710} \\
\small{E-mail: \texttt{rafael.delamadrid@lamar.edu}}}

\date{\small{March 30, 2017}}

%\date{\small{March 21, 2017}}

%\date{\small{March 2, 2017}}

%\date{\small{February 18, 2017}}

%\date{\small{January 16, 2017}}

%\date{\small{January 2, 2017}}

%\date{\small{December 20, 2016}}

%\date{\small{November 29, 2016}}

%\date{\small{October 17, 2016}}  

%\date{\small{August 24, 2016}}

%\date{\small{May 16, 2016}}

%\date{\small{December 11, 2015}}

\maketitle

\begin{abstract}
\noindent
We express the resonant energies of the delta-shell potential
in terms of the Lambert $W$ function, and we calculate their decay widths and
decay constants. The ensuing numerical results strengthen the interpretation of
such decay widths and constants as a way to quantify the coupling 
between a resonance and the continuum. We calculate 
explicitly the decay energy spectrum of the resonances of the delta-shell 
potential, and we show numerically that the lineshape of such spectrum
is not the same as, and can be very different
from, the Breit-Wigner (Lorentzian) distribution. We argue that
the standard Golden Rule cannot describe the interference of two 
resonances, and we show how to describe such interference by way of the
decay energy spectrum of two resonant states.
\end{abstract}


\newpage

\section{Introduction}
\setcounter{equation}{0}
\label{sec:intro}

Although there are several theoretical ways to describe a resonance, there is
increasing evidence in molecular, atomic, nuclear and particle physics that
a resonance should be defined as a pole of the $S$ 
matrix~\cite{TAYLOR}, due to the phenomenological and theoretical 
advantages of such definition~\cite{SIRLIN,WILLENBROCK,STUART,LEIKE,BERNICHA,
CASO,BS,ELANDER1,ELANDER3,GEGELIA,ELANDER4,CECI2,CECI1,TIATOR13,PDG,TIATOR2,
Vaandrager,TIATOR3}. Once we accept the definition of a resonance as
a pole of the $S$-matrix, it is very natural to associate with it a 
resonant (Gamow) state~\cite{GAMOW,SIEGERT,ZELDOVICH,BERGGREN,PESKIN,BOLLINI1,
TOLSTIKHIN1,MONDRAGON00,05CJP,TOLSTIKHIN2,TOLSTIKHIN3,VELAZQUEZ,NPA08,MICHEL7,
NIMROD,GASTON10,SASADA,GASTON11,GASTON12,GASTON13,FOSSEZ,ANDERSEN,JULVE14,
HATANO14,GASTONPRA14,SARA,LUNDMARK,GENTILINI,NPA15,
GASTON16,BROWN16,PLASTINO1,PLASTINO2,CEVIK,GARMON,OLENDSKI,GASTON17}. 

Because resonant states are the wave functions
of resonances, it is logical to obtain from them measurable quantities such
as decay rates and branching fractions. In Ref.~\cite{NPA15}, it was 
shown how to construct the decay energy spectrum, the differential 
and the total decay widths, and the differential and the total decay 
constants of a resonant state. In
the present paper, we will use the example of the delta-shell potential
to obtain a numerical validation of the results of Ref.~\cite{NPA15}. Such 
numerical validation is needed because there is a result of
perturbation theory~\cite{DUNCAN} that is seemingly in
conflict with the results of Ref.~\cite{NPA15}. The explicit numerical 
calculation of the quantities introduced in Ref.~\cite{NPA15} will show that
the formalism of Ref.~\cite{NPA15} is indeed sound.

Although it is not very realistic, the delta-shell 
potential~\cite{GOTTFRIED,WINTER,DICUS} is frequently used to 
exemplify and test new results in the 
theory of resonances, because it is the simplest potential that produces 
resonances, and because it is almost exactly solvable. For example, in 
Ref~\cite{MONDRAGON00}, a double
delta-shell potential was used to numerically obtain an exceptional point. In 
Ref.~\cite{GASTON11}, the delta-shell potential was utilized to study the
decay of two identical, non-interacting particles. In Ref.~\cite{GASTON12},
it was used to illustrate that the resonant states yield the same time
evolution of a wave packet as the scattering states. In
Refs.~\cite{GASTON13,GASTONPRA14}, the delta-shell
potential with a complex coupling was employed to study time evolution in
the presence of absorbing and emitting potentials. In Ref.~\cite{GASTON17},
it was used to study the non-Hermitian character of the Born rule in
open quantum systems. In 
Ref.~\cite{SCOTT}, a double well Dirac delta function model was utilized as 
the one-dimensional limit of H$_2^+$. In Ref.~\cite{SANTINI1}, the 
delta-shell potential was used to relate resonances 
with the eigenstates of a particle in a box by means of renormalization
and mixing. In Ref.~\cite{SANTINI2}, it was utilized to
study perturbative and non-perturbative dynamics of resonances. In our case,
besides its simplicity, we use the delta-shell potential because 
we have found that the transcendental equation that provides its
resonant energies can actually be solved exactly in terms of the 
Lambert $W$ function, thereby making the delta-shell potential a 
fully solvable model for resonances.

It is important to understand to what kind of situations the decay
energy spectrum of Ref.~\cite{NPA15} would
apply. Resonances appear as sharp peaks of the cross section. For example, 
in the reaction $\pi ^+ +p \to \pi ^+ +p$,
there is a sharp peak centered at 1230~MeV whose width is about 
115~MeV~\cite{SEGRE}. Such peaks are thought to be due to intermediate,
unstable particles that are described by a pole of the $S$-matrix, and they
are interpreted in terms of the cross section rather than the decay energy
spectrum of Ref.~\cite{NPA15}. Resonances are also observed 
when the energy spectrum of its decay products is
measured. For example, in the reaction $\pi ^+ +p \to \pi ^++\pi^0 +p$, we can
plot the number of decay evens of $\pi ^+\pi^0$ versus the invariant mass
of the two pions~\cite{SEGRE}. It is then found that there is a peak
around 770~MeV that corresponds to the energy of the $\rho ^+$ resonance. The
reaction is then thought to proceed in two steps. In the first one, a 
proton and the $\rho ^+$ are produced; in the second one, the $\rho ^+$ decays
into two pions. It is to this kind of situations --when the decay energy 
spectrum of a resonance is measured by counting the number of
decay products as a function of the energy-- that the decay energy 
spectrum of Ref.~\cite{NPA15} applies, in the non-relativistic domain. 

Our results are based on the assumption that the wave function of a resonance
is exactly given by a resonant state. However,
a recurring objection to the use of resonant states is that such
states are not physical. In order to counter such objection, we would
like to point out that, theoretically, using a resonant state 
to describe a resonance is complementary to, and in the same spirit as, 
using a pole of the $S$-matrix. When one describes a resonance
as the pole of the $S$-matrix (which is defined on the scattering
spectrum), one analytically continues the $S$-matrix from the real, scattering
energies (which are the only energies that are accessible by experiment), to
the complex resonant pole (which is not directly accessible by 
experiment). The Laurent expansion of the $S$-matrix is
the sum of a resonant part (the pole's contribution), and a non-resonant part 
(the background). The $S$-matrix description of a resonance assumes that the
resonance's contribution to the cross section is given by the pole's 
contribution, and
that the background has nothing to do with the resonance itself. Similarly,
when one uses a wave function $\varphi$ to describe a system with 
resonances, one analytically continues $\varphi$ from the real, scattering
energies into the complex pole of the $S$-matrix, resulting in 
an expansion of the form
$|\varphi \rangle =\Cr |\zr \rangle + |\text{bg}\rangle$. In such expansion, 
the resonant state $|\zr \rangle$ is supposed to carry
the resonance's contribution to the state $\varphi$ (including the 
exponential decay), and the background $|\text{bg}\rangle$ is supposed 
to carry the non-resonant
contributions (including deviations from exponential decay). 

Since one can use the formalism of Ref.~\cite{NPA15} to calculate 
the decay rate of a square-integrable wave function $\varphi$, one may be 
tempted to identify the decay rate of a wisely chosen $\varphi$, rather than
the decay rate of a resonant state, with the true decay 
rate of a resonance. However, describing 
the decay rate of a resonance by way of a square-integrable function
$\varphi$ leads to ambiguities. To see why, let us imagine that we are using
a wave function $\varphi _1$ to obtain the decay rate and the
lifetime of a resonance. We
can always expand $\varphi _1$ in terms of the resonant state,
$\varphi _1= \Co |{\zr} \rangle + |\text{bg}_1\rangle$. But if we used 
a wave function $\varphi _2= \Ct |\zr \rangle + |\text{bg}_2\rangle$ that is 
only slightly different from $\varphi _1$, we would obtain a slightly different
decay rate and a slightly different lifetime, and we could not tell which ones 
are the right ones, let alone relate them to the pole width. The Gamow-state 
description of resonances addresses
this ambiguity by associating a unique Gamow state to each pole of the
$S$-matrix, and by defining resonant properties such as decay rates and
lifetimes in terms of the Gamow states. In our example, there is only one 
resonance, and $\varphi _1$
and $\varphi _2$ are two different approximations of one and the same 
resonant state. The backgrounds $|\text{bg}_1\rangle$ and 
$|\text{bg}_2\rangle$ provide a measure of how well the
wave function is tuned around the resonant state. Thus, by identifying the 
resonant state with the wave function of a resonance, one has an 
unambiguous way to prescribe what is resonance from what is background.

Since we will denote similar quantities by similar symbols, it may be 
helpful to review our notation. The differential and the
total decay
{\it widths} will be denoted by $\frac{\rmd \overline{\Gamma} (E)}{\rmd E}$
and $\overline{\Gamma}$, respectively. The
differential and the total decay {\it constants} will be
denoted by $\frac{\rmd \Gamma (E)}{\rmd E}$ and $\Gamma$, respectively. The
complex, resonant energy, which is a pole of the $S$ matrix, will be 
denoted by $\zr = \er -\rmi \gr /2$, where
$\er$ is the (mean) energy of the resonant state, and $\gr$ is the
pole width. 

The structure of the paper is as follows. In Sec.~\ref{sec:review}, we 
review the main properties of the
delta-shell potential and obtain the resonant energies for the zero angular
momentum case. We will show that the transcendental equation that provides 
the resonant energies can be solved explicitly in terms of the Lambert
$W$ function. In Sec.~\ref{sec:datdw}, we calculate the decay widths
$\overline{\Gamma}$ and the decay constants $\Gamma$. We will show that
the closer the resonance is to the real axis (and hence the closer it 
is to become a bound state), the smaller $\overline{\Gamma}$ is, and the
larger $\Gamma$ is. Thus, both $\overline{\Gamma}$ and $\Gamma$ can be 
used to quantify the strength of the interaction between the resonance 
and the continuum. We will also point out that, although a standard 
result of perturbation theory seems to be in contradiction with the
formalism of Ref.~\cite{NPA15}, the numerical results of the
present paper show that such formalism is well grounded. In 
Sec.~\ref{sec:goldernue}, we use
the Golden Rule of a resonant state to obtain the values of
the decay widths and constants in the approximation that the resonance
is sharp. We will denote such approximate values 
by $\overline{\Gamma}_{\rm sharp}$ and $\Gamma _{\rm sharp}$. We will show that,
surprisingly, for truly sharp resonances $\overline{\Gamma}_{\rm sharp}$ is very
close to the pole width $\gr$, even though $\overline{\Gamma}$ is very
different from $\gr$. In Sec.~\ref{sec:spectrum}, we obtain the
theoretical probability $\frac{\rmd P (E)}{\rmd E}$ that is to correspond 
to the experimental decay energy spectrum. We will show numerically that 
the Breit-Wigner (Lorentzian) distribution does not coincide exactly with 
the natural lineshape of a resonance, even when
the effect of the threshold can be neglected. In addition, we will show that
$\frac{\rmd P (E)}{\rmd E}$ naturally suggests
a new normalization of the resonant states. We will also point
out some similarities of the decay energy spectra of the delta-shell
potential with experimental ones. In Sec.~\ref{sec:interference},
we obtain the decay energy spectrum of two interfering resonances, and we
argue that the standard Golden Rule is not suitable to describe such
interference. Although cross sections are not the main focus of the
present paper, in 
Sec.~\ref{sec:crosspeaks} we compare three different approximations
of the $S$-matrix to describe resonant peaks in cross sections.
Finally, in Sec.~\ref{sec:conclusions} we state 
our conclusions.

\section{The delta-shell potential}
\setcounter{equation}{0}
\label{sec:review}

Let us review the main features of the delta-shell 
potential~\cite{GOTTFRIED}.
A delta-shell potential at $r=a$ is given by
\begin{equation}
       V(r)=g\delta (r-a) \, ,
	\label{deltap}
\end{equation}
where $g$ characterizes the opaqueness of the barrier. In the
radial, position representation, and for zero angular momentum, 
the eigensolutions of the time-independent Schr\"odinger equation 
subject to appropriate boundary conditions are given by
\begin{equation}
       \chi (r;E)=A(E) \left\{ \begin{array}{ll}
        \sin (kr) \quad  & 0<r<a \, , \\ [2ex]
        {\cal J}_1(E) \rme ^{\rmi kr} + {\cal J}_2(E) \rme ^{-\rmi kr}        
     \quad &  a<r<\infty \, , 
        \end{array} 
      \right. 
	\label{0-}
\end{equation}
where $k=\sqrt{\frac{2m}{\hbar ^2}E}$ is the wave number,
$A(E)$ is a normalization constant, and the Jost functions are given by
\begin{equation}
       {\cal J}_{{\,}_{2}^{1}}(E)   = \frac{1}{4k} \left[ \mp 
                       2\rmi k + \frac{2mg}{\hbar ^2} 
                        \left( \rme ^{\mp 2\rmi ka} -1\right) \right] \, .
               \label{j1a2}  
\end{equation}
The $S$-matrix is given by
\begin{equation}
      S(E)=-\frac{ {\cal J}_1(E)}{{\cal J}_2(E)}  \, .
\end{equation}
The resonant wave numbers are given by the poles of the $S$-matrix, 
which coincide with the zeros of ${\cal J}_2(E)$,
\begin{equation}
       2\rmi k a + \lambda 
                        \left( \rme ^{2\rmi k a} -1\right) =0 \, ,
       \label{complexroots}
\end{equation}
where $\lambda = \frac{2m}{\hbar ^2}ga$ is a dimensionless
constant that characterizes the strength of the 
potential. Equation~(\ref{complexroots}) is a transcendental equation
that has an infinite number of complex solutions. Such solutions
can be written in terms of the Lambert $W$ 
function~\cite{LAMBERT1,LAMBERT2,LAMBERT3,LAMBERT4,LAMBERT5,LAMBERT6,
NOTE}. The 
resonant wave numbers, which lie in the fourth quadrant, are given by
\begin{equation}
     k_n\equiv \alpha _n-\rmi \beta _n=
\left\{ \begin{array}{lc}
         \frac{1}{2\rmi a } \left[ \lambda -
              W_{-n}(\lambda  \rme ^{\lambda }) \right] \, , \qquad
            &  n = 1, 2, \ldots ,   \quad \lambda > 0 \, , \\ [2ex]
         \frac{1}{2\rmi a } \left[ \lambda -
              W_{-n}(\lambda  \rme ^{\lambda }) \right] \, , \qquad
            &  n = 2, 3, \ldots ,  \quad  \lambda < 0 \, , 
     \end{array}
            \right.
         \label{complexsol}
\end{equation}
where $W_n$ is the $n$th branch of the Lambert $W$ function. For 
$\lambda <-1$, the delta-shell potential forms a bound state whose
wave number is given by
\begin{equation}
     k_{\rm b}\equiv \rmi |k_{\rm b}|= \frac{1}{2\rmi a } \left[ \lambda -
              W_{0}(\lambda  \rme ^{\lambda }) \right] \, .
        \label{boundstate}
\end{equation}
For $-1< \lambda <0$, the delta-shell potential forms a virtual (also
known as anti-bound) state of wave number
\begin{equation}
     k_{\rm v}\equiv -\rmi |k_{\rm v}|= \frac{1}{2\rmi a } \left[ \lambda -
              W_{-1}(\lambda  \rme ^{\lambda }) \right] \, .
        \label{virtualstate}
\end{equation}
For any $\lambda$, the 
anti-resonant wave numbers, which lie in the third quadrant, are given by
\begin{equation}
          k_{-n}\equiv -\alpha _n-\rmi \beta _n= -k_n^* =
              \frac{1}{2\rmi a } \left[ \lambda  -
              W_{n}(\lambda  \rme ^{\lambda }) \right] \, , \qquad
             n = 1, 2, \ldots \, .
        \label{complexsolar}
\end{equation}

By using
a program such as Mathematica, Maple or Matlab, it is very easy to 
calculate the resonant wave numbers of the delta-shell potential for a given 
value of $\lambda$. Tables~\ref{table:100}-\ref{table:-100}
in Appendix~\ref{app:tables} list the first 
eight resonant wave numbers and energies for the cases 
$\lambda = \pm 0.5, \pm 10,$ and $\pm 100$ in units 
where $a=\hbar ^2/2m=1$. This choice of units is equivalent to measuring
wave number in units of $1/a$ and energy in units of $\hbar ^2/2ma^2$. The
virtual ($\lambda = -0.5$, Table~\ref{table:-0.5}) and
bound ($\lambda = -10$, Table~\ref{table:-10}; $\lambda = -100$, 
Table~\ref{table:-100}) states are also listed.

The resonant eigenfunctions are given by
\begin{equation}
       \langle r|\zr \rangle = u(r;\zr)= \nr \left\{ \begin{array}{ll}
        \frac{1}{{\cal J}_1(\zr)} \sin (\kr r) \quad  & 0<r<a \, , \\ [2ex]
                  \rme ^{\rmi \kr r}        
     \quad &  a<r<\infty \, ,
        \end{array} 
      \right. 
	\label{rstate}
\end{equation}
where Zeldovich's normalization constant $\nr$ is given by the
residue of the $S$-matrix at the complex resonant wave number $\kr$, 
\begin{equation}
       \nr ^2= \rmi \, \text{res}\left[S(q)\right]_{q=\kr} =
              -\rmi \, \frac{{\cal J}_1(\kr)}{{\cal J}_2'(\kr)} \, .
\end{equation}

\section{The decay width and the decay constants}
\setcounter{equation}{0}
\label{sec:datdw}

In Ref.~\cite{NPA15}, we identified 
$|\langle E|\rme ^{-\rmi H\tau /\hbar}|\zr \rangle|^{2}$ with
the probability density $\frac{\rmd p_{\tau}}{\rmd E}$ 
that the resonance has decayed into a stable particle of energy $E$ at
time $\tau$,
\begin{equation}
      \frac{\rmd p_{\tau}}{\rmd E}\equiv 
         |\langle E|\rme ^{-\rmi H\tau /\hbar}|\zr \rangle|^{2}  \, .
             \label{detdpd}
\end{equation}
The differential decay width associated with such a probability is defined
as (see Refs.~\cite{ANDREASSEN1,ANDREASSEN2} for a somewhat related definition)
\begin{equation}
      \frac{\rmd \overline{\Gamma} (E)}{\rmd E}  \equiv
         -\hbar \frac{\rmd}{\rmd \tau}\left(
               \left.\frac{\rmd  p_{\tau}}{\rmd E}\right)
                   \right|_{\tau =0}    \, .
             \label{diffdr}
\end{equation}
As shown in Ref.~\cite{NPA15}, the differential and the total decay widths of a 
resonant state can be written as
\begin{eqnarray}
   \frac{\rmd \overline{\Gamma} (E)}{\rmd E}=
       \frac{\gr}{(E-\er)^2+(\gr/2)^2} |\langle E|V \gs|^2 \, , 
         \label{inteGam6n}  \\
           \overline{\Gamma} = \int_0^{\infty} \rmd E \
       \frac{\gr}{(E-\er)^2+(\gr/2)^2} |\langle E|V \gs|^2 \, . 
         \label{inteGam6totaln}
\end{eqnarray}
The decay width $\overline{\Gamma}$ has units of energy, and physically
can be interpreted as the initial decay rate associated with 
the probability $p_{\tau}$. Thus, at least in principle, one could measure
$\overline{\Gamma}$ by measuring the initial decay rate of 
$p_{\tau}$. However, in general $\overline{\Gamma}$ is different from $\gr$.
Thus, $\overline{\Gamma}$ is not the same as the width of resonant
peaks in cross sections, because such width is determined by $\gr$.

The differential and the total decay constants of a resonant state can be 
written as~\cite{NPA15}
\begin{eqnarray}
   \frac{\rmd \Gamma (E)}{\rmd E} =
       \frac{1}{(E-\er)^2+(\gr/2)^2} |\langle E|V \gs|^2 \, , 
         \label{inteGam6ti} \\ [2ex]
           \Gamma =  \int_0^{\infty} \rmd E \
       \frac{1}{(E-\er)^2+(\gr/2)^2} |\langle E|V \gs|^2 \, . 
         \label{inteGam6totalti}
\end{eqnarray}
The differential decay constant $\frac{\rmd \Gamma (E)}{\rmd E}$ 
has units of $\left({\rm energy}\right)^{-1}$ and hence can be
interpreted as the probability density that the resonance decays into
a stable particle of energy $E$. By contrast,
$\Gamma$ is a dimensionless constant that is (formally) equal to
the norm of the absolute value squared of the resonant 
state~\cite{NPA15}. Thus, at first sight, it may seem that $\Gamma$ 
can be interpreted as
the total probability of transition from a resonant state to a 
continuum. However, because it can be greater than unity, $\Gamma$
doesn't have a straightforward probabilistic interpretation. This is
actually a general feature of resonant states: Some quantities that
for bound and scattering states have a straightforward probabilistic 
interpretation become greater than unity for resonant states (see
for example Ref.~\cite{GASTON17}). It was proposed in Ref.~\cite{GASTON17}
that such quantities may have an interpretation as ``quasiprobabilities,''
although it is still an open question whether $\Gamma$ can be interpreted 
that way. In the present paper, $\Gamma$ will be absorbed in the definition of
the decay energy spectrum (see
Sec.~\ref{sec:spectrum}), and therefore it will not matter that $\Gamma$ can
be greater than unity.

From Eqs.~(\ref{inteGam6n})-(\ref{inteGam6totalti}), there follows that
\begin{eqnarray}
       & \frac{\rmd \Gamma (E)}{\rmd E}= \frac{1}{\gr} 
             \frac{\rmd \overline{\Gamma} (E)}{\rmd E} \, , 
            \label{rddcddw}  \\ [1ex]
      & \Gamma =  \frac{\overline{\Gamma}}{\gr} \, .
             \label{redcdw}
\end{eqnarray}

For the delta-shell potential, the differential and the total decay widths
are given by~\cite{NPA15}
\begin{equation}
   \frac{\rmd \overline{\Gamma} (E)}{\rmd E}=\frac{1}{\pi}
       \frac{\gr/2}{(E-\er)^2+(\gr /2)^2} 
        \frac{4mg^2}{\hbar ^2} \,
      \frac{\sin ^2\left(k a \right)}{k} |\nr|^2 \rme^{2\br a}    \, , 
         \label{inteGam6dp}
\end{equation}
\begin{equation}
   \overline{\Gamma}  =  \frac{4mg^2}{\hbar ^2} |\nr|^2 \rme ^{2\br a} C 
         = \frac{\hbar ^2}{2ma^2} 2 \lambda ^2  |\nr|^2 \rme ^{2\br a} C \, ,
     \label{inteGam7dpb}
\end{equation}
where
\begin{equation}
   C = \int_0^{\infty}\rmd E \, \frac{1}{\pi}\frac{\gr/2}{(E-\er)^2+(\gr /2)^2} 
        \frac{\sin ^2\left(k a \right)}{k}  \, .
        \label{constantc}
\end{equation}
Tables~\ref{table:100}-\ref{table:-100} in Appendix~\ref{app:tables} 
list the values of 
$\overline{\Gamma}$ for $\lambda = \pm 0.5, \pm 10$, and $\pm 100$ 
in units of $\hbar /2m a^2$ for the first eight resonances of the
delta-shell potential. The values of $\Gamma$, obtained through 
Eq.~(\ref{redcdw}), are also listed. It follows from 
Tables~\ref{table:100}-\ref{table:-100} that the decay
width $\overline{\Gamma}$ is different from the $S$-matrix width
$\gr$ (if $\overline{\Gamma}$ and $\gr$ were equal,
by Eq.~(\ref{redcdw}) $\Gamma$ would always be equal to 1). We can see in 
Tables~\ref{table:100}-\ref{table:-100} that, for a given value of 
$\lambda$, the closer the resonance is to the real axis, the 
larger the decay constant $\Gamma$ is, and the smaller the decay width 
$\overline{\Gamma}$ is. We can also see 
in Tables~\ref{table:100}-\ref{table:-100} that if we follow a given
resonance, say the first one, for different values of $\lambda$, then
as the magnitude of $\lambda$ increases and therefore the resonance
becomes more stable, the decay width $\overline{\Gamma}$ becomes smaller and 
the decay constant $\Gamma$ becomes larger. Thus, $\overline{\Gamma}$
and $\Gamma$ can be seen as parameters that quantify how strongly the 
resonance couples to the continuum.

As can be seen in Tables~\ref{table:-10} and~\ref{table:-100}, for a bound 
state $\overline{\Gamma}$ is equal to 0, as you would expect from a 
stable state. In 
addition, for a bound state the decay constant is equal to 1. The reason
is that, as can be seen from its definition~\cite{NPA15}, the
decay constant of a bound state $|E_{\rm bound}\rangle$ is equal to the 
norm squared of the bound-state wave function,
\begin{equation}
      \Gamma = \int_0^{\infty} 
    |\langle E|E_{\rm bound}\rangle |^2 \rmd E = 
       \langle E_{\rm bound}|E_{\rm bound}\rangle = 
      \| |E_{\rm bound} \rangle \| ^2 \, .
\end{equation}
Since for bound states Zeldovich's normalization coincides with the usual 
normalization of the absolute value squared of 
the wave function, we have that $\| |E_{\rm bound} \rangle \| ^2=1$,
and therefore $\Gamma$ must be equal to 1 for bound states. 

Similar to a bound state, for a virtual state 
$\overline{\Gamma}$ is also equal to 0 (see Table~\ref{table:-0.5}). However,
unlike for a bound state, the value of $\Gamma$ for a virtual state is
not equal to 1.

Overall, the numerical results of Tables~\ref{table:100}-\ref{table:-100}
provide a numerical validation of the formalism of Ref.~\cite{NPA15}. Such
validation was needed because
there is a result of perturbation theory that is in contradiction
with the formalism of Ref.~\cite{NPA15}. As can be seen for example
on page 200 of Ref.~\cite{DUNCAN}, {\it second-order} perturbation theory can
be used to show that the pole width for the transition from 
an initial state $|i\rangle$ of complex energy $E_i+\delta -\rmi \, \gr/2$
to a set of final states $|n\rangle$ satisfies the following implicit
equation:
\begin{equation}
        \gr = \sum_{n\neq i} |V_{ni}|^2 
           \frac{\gr}{(E_i+\delta -E_n)^2+ (\gr /2)^2} \, ,
       \label{duncan1}
\end{equation}
where $E_i$ is the energy of the initial state, $\delta$ is the energy
shift, and $V_{ni}$ is the matrix element between
$|i\rangle$ and $|n\rangle$. The continuum version of 
Eq.~(\ref{duncan1}) can be written as
\begin{equation}
        \gr = \int_0^\infty \rmd E \, 
           \frac{\gr}{(\er -E)^2+ (\gr /2)^2} |\langle E|V|i\rangle|^2  \, .
       \label{duncan2}
\end{equation}
If we take $|i\rangle$ to be a Gamow state $|\zr\rangle$, then 
Eqs.~(\ref{inteGam6totaln}), (\ref{redcdw}) and~(\ref{duncan2}) 
would imply that 
the total decay width $\overline{\Gamma}$ is equal to the pole width
$\gr$, and that the decay constant $\Gamma$ is always 1,
\begin{equation}
       \overline{\Gamma} =\gr \, ,  \hskip1cm \Gamma =1 \, .
         \label{pt}
\end{equation}
This would render the
results of Ref.~\cite{NPA15} invalid, since in appendix~A of Ref.~\cite{NPA15}
it was shown that $\overline{\Gamma}$ and $\gr$ are different,
\begin{equation}
       \overline{\Gamma} \neq \gr \, ,  \hskip1cm \Gamma \neq 1 \, .
         \label{us}
\end{equation}
Obviously, Eqs.~(\ref{pt}) and~(\ref{us}) cannot both be correct. Since our 
numerical results show that Eq.~(\ref{us}), rather than
Eq.~(\ref{pt}), is correct, we can conclude that the formalism of 
Ref.~\cite{NPA15} is sound. 

In principle, perturbation theory and the results of Ref.~\cite{NPA15} should
agree with each other, since $\gr$ is determined by the pole
of the $S$-matrix, not by perturbation theory or by
the formalism of Ref.~\cite{NPA15}. It is still an open question why the 
formula in Eq.~(\ref{duncan2}) is not correct.

The formalism of Ref.~\cite{NPA15} can also be used to define partial widths
and branching fractions for a resonance that has more than one mode of
decay, and we would like to compare them with the partial widths of
Refs.~\cite{PESKIN,GASTON10}. Like Refs.~\cite{PESKIN,GASTON10},
Ref.~\cite{NPA15} uses the resonant states to define the
partial widths. However, the partial widths of Refs.~\cite{PESKIN,GASTON10}
use only the tails of the resonant states, and they are more appropriate
to describe tunneling of a particle through a potential barrier, whereas 
the partial widths of Ref.~\cite{NPA15} use the whole resonant wave
function and describe decay of a resonance into the continuum. Thus, both
approaches describe different, although related, resonance phenomena. In 
addition, the formalism of Ref.~\cite{NPA15} allows us to define branching
fractions as the ratio of two dimensionless quantities (as it is done
in experiments), whereas the branching fractions of 
Refs.~\cite{PESKIN,GASTON10} would be given by the ratio of two dimensionful
quantities.

\section{The Golden Rule of a resonant state}
\setcounter{equation}{0}
\label{sec:goldernue}

As shown in Ref.~\cite{NPA15}, by (formally) replacing the Lorentzian
by the delta function when the resonance is sharp, we can obtain the
Golden Rule of a resonant state,
\begin{eqnarray}
    && \frac{\rmd \overline{\Gamma} (E)}{\rmd E}\approx
       \frac{\rmd \overline{\Gamma}_{\rm sharp} (E)}{\rmd E} =
 2\pi 
         |\langle E|V \gs|^2 \delta (E-\er) \, , 
         \label{inteGam8} \\ [2ex]
    && \overline{\Gamma} \approx   \overline{\Gamma}_{\rm sharp}=
          2\pi  \,  
         |\langle \er |V \gs|^2  \, . 
         \label{inteGam9}
\end{eqnarray}
In addition, because of Eq.~(\ref{redcdw}), the 
decay constant of a sharp resonance can be approximated by
\begin{equation}
       \Gamma \approx 
      \Gamma _{\rm sharp}= \frac{\overline{\Gamma}_{\rm sharp}}{\gr} \, .
        \label{gammasharp}
\end{equation}

For the delta-shell potential, Eqs.~(\ref{inteGam8}) and~(\ref{inteGam9})
can be written as
\begin{equation}
    \frac{\rmd \overline{\Gamma}_{\rm sharp} (E)}{\rmd E} =
         \frac{4mg^2}{\hbar ^2} \,
      \frac{\sin ^2\left(k a \right)}{k} |\nr|^2 \rme^{2\br a}   \delta (E-\er) 
                \, , 
         \label{inteGam10}
\end{equation}
\begin{equation}
    \overline{\Gamma}_{\rm sharp}   = 
\frac{4mg^2}{\hbar ^2} \,
      \frac{\sin ^2 (\tilde{k}_{\text{\tiny R}} a )}{\tilde{k}_{\text{\tiny R}}} 
        |\nr|^2 \rme^{2\br a} 
       =  \frac{\hbar ^2}{2ma^2} 2 \lambda ^2
      \frac{\sin ^2 (\tilde{k}_{\text{\tiny R}} a )}{\tilde{k}_{\text{\tiny R}}} 
        |\nr|^2 \rme^{2\br a} \, ,
         \label{inteGam11b} 
\end{equation}
where $\tilde{k}_{\text{\tiny R}}=\sqrt{(2m/\hbar ^2)\er}$. 
Tables~\ref{table:100}-\ref{table:-100} in Appendix~\ref{app:tables} list 
the values of 
$\overline{\Gamma}_{\rm sharp}$ when $\lambda = \pm 0.5, \pm 10$ 
and $\pm 100$ in units of $\hbar ^2 /2m a^2$ for the first eight 
resonances of the delta-shell potential. The values of $\Gamma _{\rm sharp}$ 
are obtained through Eq.~(\ref{gammasharp}). Surprisingly, for sharp 
resonances the value of 
$\overline{\Gamma}_{\rm sharp}$ is very close to the value of 
$\gr$, even though the exact value of the
decay constant $\overline{\Gamma}$ is not
close to $\gr$. The closeness of 
$\overline{\Gamma}_{\rm sharp}$ to $\gr$ is also manifest
in the value of $\Gamma _{\rm sharp}$: As the resonance becomes sharper, 
$\Gamma _{\rm sharp}$ becomes closer to 1.

That $\overline{\Gamma}_{\rm sharp}$ is close to the pole width $\gr$ 
for sharp resonances (or,
equivalently, that $\Gamma _{\rm sharp}$ is close to 1)
is not very surprising, since for sharp resonances the standard Golden Rule 
should yield the pole width $\gr$. What is surprising is that the exact value
of the decay width $\overline{\Gamma}$ is not approximately the same as
the pole width $\gr$, even for very sharp resonances. Numerically, the reason 
is that the exact value of the integral in Eq.~(\ref{constantc}) is not the 
same as what one obtains by formally replacing the Lorentzian by the 
delta function, even when $\gr \approx 10^{-3}$. Theoretically, this
means that although the results of Ref.~\cite{NPA15} agree with Fermi's 
Golden Rule (which is a result of {\it first-order} perturbation theory)
when we can replace the Lorentzian by the delta function, the formalism 
of Ref.~\cite{NPA15}
disagrees with a result of {\it second-order}
perturbation theory, as explained in Sec.~\ref{sec:datdw}.

\section{The decay energy spectrum}
\setcounter{equation}{0}
\label{sec:spectrum}

The s-wave partial cross section $\sigma (E)$ is given by~\cite{TAYLOR}
\begin{equation}
      \sigma (E)= \frac{\pi}{k^2} |S(E)-1|^2 \, .
       \label{crossec}
\end{equation}
When $S(E)$ has a pole at the complex energy $\zr$, we can expand $S(E)-1$ in
a Laurent expansion in a region close to $\zr$,
\begin{equation}
       S(E)-1= \frac{{\rm r}_{\text{\tiny R}}}{E-\zr} +B(E) \, ,
       \label{laurent}
\end{equation}
where ${\rm r}_{\text{\tiny R}}$ is the residue of $S(E)$ at $\zr$ and $B(E)$
is an analytic function that corresponds to the background. By identifying
the pole's contribution to the $S$-matrix with the resonance's contribution
to the cross section, and by neglecting the background, we approximate the
cross section in the vicinity of the resonant energy as follows,
\begin{equation}
      \sigma (E)\approx \sigma_{\text{\tiny Laurent}}(E) =\frac{\pi}{k^2} 
\frac{|{\rm r}_{\text{\tiny R}}|^2}{(E-\er)^2+(\gr/2)^2} \, ,
        \label{arppc}
\end{equation}
which is the well-known Breit-Wigner formula. The approximation in
Eq.~(\ref{arppc}) is valid when the resonance is isolated and far from 
a threshold~\cite{NOTE2}. We then say that the Lorentzian
peak in the cross section is produced by an intermediate, unstable particle
of (mean) energy $\er$, width $\gr$, and lifetime 
$\tau _{\text{\tiny R}} =\frac{\hbar}{\gr}$.

When one measures the number of decay events per energy bin, instead of the 
cross section one obtains the decay energy spectrum. If the 
number of decay events per energy bin is normalized by
the total number of events, the resulting normalized decay energy spectrum
is interpreted as the probability density that the
resonance decays into the continuum. In Ref.~\cite{NPA15}, it was proposed 
that the differential decay constant $\frac{\rmd \Gamma (E)}{\rmd E}$ 
describes such probability density. However, since 
$\Gamma = \int_0^{\infty}\frac{\rmd \Gamma (E)}{\rmd E} \rmd E \neq 1$, the
probability distribution $\frac{\rmd \Gamma (E)}{\rmd E}$ is not normalized 
to $1$. We therefore define the probability density
\begin{equation}
     \frac{\rmd P (E)}{\rmd E} \equiv
   \frac{1}{\Gamma} \frac{\rmd \Gamma (E)}{\rmd E}=\frac{1}{\Gamma}
    \frac{1}{(E-\er)^2+(\gr/2)^2} |\langle E|V \gs|^2  \, .
        \label{defP}
\end{equation}
Clearly, $\frac{\rmd P (E)}{\rmd E}$ is normalized to $1$. Thus,
we can interpret the theoretical probability distribution 
$\frac{\rmd P (E)}{\rmd E}$ as the normalized experimental decay 
energy spectrum.

It should be noted that the
reason why $\frac{\rmd \Gamma (E)}{\rmd E}$ is not normalized to $1$ is that
Zeldovich's normalization factor $\nr$ ensures that the square of the resonant
state, rather than its absolute value squared, is normalized to $1$. We can
however normalize the resonant states as 
$\frac{1}{\sqrt{\Gamma}}|\zr \rangle$, 
where $|\zr \rangle$ is the resonant state normalized according to
Zeldovich's prescription. The advantage of such normalization is that
the resonant state $\frac{1}{\sqrt{\Gamma}}|\zr \rangle$ automatically 
yields a probability distribution $\frac{\rmd P (E)}{\rmd E}$ that is 
normalized to $1$.

For the delta-shell potential, it follows from Eqs.~(\ref{inteGam6ti})
and~(\ref{defP}) that 
\begin{equation}
        \frac{\rmd P (E)}{\rmd E} = \frac{1}{2\pi\Gamma}
       \frac{1}{(E-\er)^2+(\gr /2)^2} 
        \frac{4mg^2}{\hbar ^2} \,
      \frac{\sin ^2\left(k a \right)}{k} |\nr|^2 \rme^{2\br a}    \, .
      \label{esd}
\end{equation}
In Fig.~\ref{fig:thirdthree100} of Appendix~\ref{app:plots}, we plot 
the decay energy spectrum of
Eq.~(\ref{esd}) along with the (normalized) Breit-Wigner distribution
$\frac{1}{\pi}\frac{\gr/2}{(E-\er)^2+(\gr /2)^2}$
and the matrix element of the interaction $|\langle E|V|\zr \rangle |^2$ for
the third resonance of the delta-shell potential when
$\lambda =100$. For the sake of clarity, Fig.~\ref{fig:thirdtwo100} 
of Appendix~\ref{app:plots} contains
the same plots as Fig.~\ref{fig:thirdthree100} except for the matrix 
element. Figure~\ref{fig:thirdtwo10} displays the decay energy spectrum 
and the Breit-Wigner distribution of the third resonance
when $\lambda =10$.

Figures~\ref{fig:thirdthree100}-\ref{fig:thirdtwo10} exhibit some features
that are common to all the resonances produced by the 
delta-shell potential. First, although the lineshape of the resonant cross 
section~(\ref{arppc}) is given by the Lorentzian, the lineshape of the
decay energy spectrum is not just the Breit-Wigner 
(Lorentzian) distribution. Rather, it is the Breit-Wigner distribution 
modulated by the matrix element. Second, the third resonance of the
delta-shell potential is very sharp when $\lambda =100$, and in this case 
the decay energy spectrum is fairly symmetric and has a shape that is
similar to the Breit-Wigner 
distribution (see Figs.~\ref{fig:thirdthree100} 
and~\ref{fig:thirdtwo100}). However, 
when $\lambda =10$, the third resonance is less sharp and the peak is 
more asymmetric (see Fig.~\ref{fig:thirdtwo10}). Although most experimental 
resonant peaks are symmetric, one can also find asymmetric ones, see 
for example figure~1 in Ref.~\cite{BESIII} and figure~2b in
Ref.~\cite{DETTORI}. In addition, when
the resonance is very sharp, the tails of the decay energy spectrum tend
smoothly to zero in regions away from the resonance's position 
(see Figs.~\ref{fig:thirdthree100} and~\ref{fig:thirdtwo100}), whereas 
the tails of a resonance that is not so sharp have small wiggles 
(see Fig.~\ref{fig:thirdtwo10}). It is actually not unusual that 
decay spectra exhibit small wiggles, as can be seen for example
in figure~1 of Ref.~\cite{LHCb1}, figure~1b of Ref.~\cite{LHCb2},
figure~1 of Ref.~\cite{LHCb3}, and figure~2 of Ref.~\cite{LHCb4}, although 
the wiggles of the figures of
Refs.~\cite{LHCb1,LHCb2,LHCb3,LHCb4} may be just statistical 
fluctuations. Third, fits using 
only the Breit-Wigner distribution would yield different
resonant parameters than fits using the exact lineshape of
Eq.~(\ref{esd}). Fourth, it is clear from 
Fig.~\ref{fig:thirdthree100} that the matrix element is very different 
from either the Breit-Wigner distribution or from the 
resonant lineshape. Thus, the matrix element should not be used as the
resonant lineshape. Fifth, Eq.~(\ref{defP}) takes 
threshold effects into account automatically. In fact, one can see
in Fig.~\ref{fig:thirdtwo10} that there is a small enhancement of the decay
energy spectrum at the $E=0$ threshold. Sixth, many resonant 
bumps are fitted by way of a
Breit-Wigner distribution with an energy-dependent width. The resonant 
lineshape of Eq.~(\ref{defP}) has an energy dependence carried by the 
matrix element, and 
therefore it may provide a useful alternative to Breit-Wigner distributions
with an energy-dependent width. Seventh, experimental resonant peaks
rarely have a Lorentzian shape, partly due to detector resolution. Often,
when the pole width of the resonance is very small compared
to the detector resolution, the width of the experimental peak is dominated 
by detector resolution, and the data are fit with 
a Gaussian distribution; when the pole width is comparable to the 
detector resolution, the data are fit with a Breit-Wigner distribution
(for the resonant lineshape) convoluted by a Gaussian distribution
(for the detector resolution). In fact, only 
in M\"ossbauer-like experiments one 
obtains peaks with Lorentzian shape. Thus, it is unlikely that the resonant 
lineshape of Eq.~(\ref{defP}) shows up directly in most experiments, unless 
one makes use of a M\"ossbauer-like effect. Eighth, the resonant 
states are exponentially-growing, non-normalizable states, although 
they can be normalized using Zeldovich's regulator. Surprisingly, though, 
they yield in a natural way a probability distribution that is finite 
without the need of any regulator. Ninth, it follows from 
Eqs.~(\ref{rddcddw}), (\ref{redcdw}) and~(\ref{defP}) that
\begin{equation}
     \frac{\rmd P (E)}{\rmd E}=
\frac{1}{\, \overline{\Gamma}\, } \frac{\rmd \overline{\Gamma} (E)}{\rmd E} \, .
        \label{defPdc}
\end{equation}
Hence, one can also use the decay width to obtain the decay energy spectrum of 
a resonant state. Tenth, similar to a Fano lineshape, the decay energy 
spectrum is asymmetric (although such asymmetry is barely noticeable when
the resonance is sharp). However,
whereas a Fano resonance appears in the cross section due to the interference
between a scattering resonance and a background, the asymmetry of the decay 
energy spectrum is an intrinsic property of the decay of a resonance into
the continuum. 

The decay energy spectra of the other resonances of the delta-shell potential
are qualitatively the same as those shown in 
Figs.~\ref{fig:thirdthree100}-\ref{fig:thirdtwo10}. However, 
for a virtual state, the decay energy spectrum has a different
shape. Figure~\ref{fig:esdvirtual} shows
the lineshape of the virtual state of the delta-shell potential when
$\lambda =-0.5$. As can be seen in Fig.~\ref{fig:esdvirtual}, the decay
energy spectrum of a virtual state is simply a sharp peak
located at the $E=0$ threshold. The peak has no resemblance with a 
Lorentzian. Rather, it is just a sharp enhancement of probability at the
threshold, which is also how virtual states appear in cross sections.

In many systems, not just one but several resonances are 
excited. Figure~\ref{fig:spikes} shows the decay energy spectrum of the
first three resonances of the delta-shell potential when 
$\lambda =100$. Because each decay spectrum is normalized to 1, you can
conclude from Fig.~\ref{fig:spikes} that the peak of the
first resonance is much sharper than the peak of the second resonance,
which in turn is much sharper than the peak of the third resonance. This
is a general feature: The higher the order of the resonance, the less 
sharp it is. In addition, as $|\lambda|$ decreases, the resonances become
less sharp.

The natural question is, what kind of experiments would this resonant
lineshape apply to? Let us imagine (for the sake of argument) that we 
have eight neutrons, six protons, and six electrons, and that we somehow have 
a procedure to produce a $^{14}$C atom out of them. Our experimental 
procedure would produce a state described by a square-integrable wave 
function $\varphi$. Assuming for simplicity that our system has
only one resonance $\zr$, we can expand $\varphi$ as
$\varphi = \Cr |\zr \rangle + |\text{bg}\rangle$. When one 
devises an experimental procedure to produce a
wave function very sharply tuned around the resonant state, the background
term is nearly zero. However, $|\text{bg}\rangle$ can never be
exactly zero, which reflects the experimental impossibility that a given
preparation procedure always yields a resonant state. In the example 
of $^{14}$C atoms, this means that no matter what our procedure is to
handle the protons, electrons and neutrons, we cannot always produce 
a $^{14}$C atom, not even in principle. During some trials, we will produce 
states that are not $^{14}$C atoms, and we identify such non-resonant
states with the background. However, in those trials where we do 
produce a $^{14}$C atom, we let it decay. By
repeating the process many times, and by discarding the trials where
we do not produce a $^{14}$C atom,  we measure
the decay energy spectrum. Such spectrum should correspond to
$\frac{\rmd P (E)}{\rmd E}$, because the wave function of the $^{14}$C atom
would be a resonant state. Thus, in any experiment where we are able
to create an unstable state that subsequently decays on its own, the
decay energy spectrum would be given by $\frac{\rmd P (E)}{\rmd E}$.

\section{Interference of two resonances}
\setcounter{equation}{0}
\label{sec:interference}

There are many instances where two (or more) resonances interfere 
because they are not isolated from each other. The $S$-matrix description
of such interference is as follows. Instead of the Laurent expansion
of Eq.~(\ref{laurent}), one uses a Mittag-Leffler expansion. For our
purposes, the Mittag-Leffler expansion can be easily obtained by means
of Cauchy's residue theorem, as done for example in 
Ref.~\cite{ELANDER1}. Let us assume 
that $S(E)$ has two poles (a higher number of interfering resonances can 
be handled analogously) $z_1$ and $z_2$ in a region enclosed by a contour
$\gamma$. The contour $\gamma$ also encloses a portion of the real axis, 
including the scattering energy $E$. Then Cauchy's residue theorem implies that
\begin{equation}
        \oint_{\gamma} \frac{S(w)-1}{w-E} \rmd w =
       2\pi \rmi \left( S(E)-1 \right) 
        + 2\pi \rmi \left( \frac{{\rm r}_1}{z_1-E} +
           \frac{{\rm r}_2}{z_2-E}    \right) \, , 
         \label{cauthe}
\end{equation}
where ${\rm r}_1$ and ${\rm r}_2$ are the residues of the $S$-matrix at 
$z_1$ and $z_2$, respectively. Equation~(\ref{cauthe}) leads to the following
Mittag-Leffler expansion:
\begin{equation}
       S(E)-1= \left( \frac{{\rm r}_1}{E-z_1} +
           \frac{{\rm r}_2}{E-z_2} \right) + \frac{1}{2\pi \rmi}
         \oint_{\gamma} \frac{S(w)-1}{w-E} \rmd w \equiv
        \left( \frac{{\rm r}_1}{E-z_1} +
           \frac{{\rm r}_2}{E-z_2} \right) + B(E) \, ,
         \label{cauthe2}
\end{equation}
where $B(E)$ is the background. By neglecting the background
in the expansion~(\ref{cauthe2}), and by substituting the result into 
Eq.~(\ref{crossec}), we obtain that
\begin{equation}
      \sigma (E)\approx \frac{\pi}{k^2}
     \left[ 
\frac{|{\rm r}_1|^2}{(E-E_1)^2+(\Gamma _1 /2)^2}
+ \frac{|{\rm r}_2|^2}{(E-E_2)^2+(\Gamma _2 /2)^2}
+2\, \text{Re} \left( \frac{{\rm r}_1{\rm r}_2^*}{(E-z_1)(E-z_2^*)}\right)
\right] .
       \label{crossecinter}
\end{equation}
Thus, the contribution to the cross section of two interference resonances
consists of two Lorentzians (these are the contributions from
each individual resonance) and an interference term.

When one measures the decay energy spectrum of two 
interfering resonances (see for example Ref.~\cite{BESIII2}),
the quantity to consider is $\frac{\rmd P (E)}{\rmd E}$ rather than
the cross section. As pointed out
in Ref.~\cite{NPA15}, one can define the differential decay energy spectrum of
any square-integrable function as
\begin{equation}
       \frac{\rmd P(E)}{\rmd E} \equiv |\langle E|\varphi \rangle |^2 \, .
      \label{ddconsv}
\end{equation}
When we describe a system of two interfering resonances, in analogy 
to Eq.~(\ref{cauthe2}), we
expand $\varphi$ in terms of the corresponding resonant states as
\begin{equation}
 |\varphi \rangle = c_1 |z_1\rangle + c_2 |z_2\rangle + |\text{bg}\rangle \, .
\end{equation}
By neglecting the background $|\text{bg}\rangle$, we extract the 
contribution to $\varphi$ of those two resonances,
\begin{equation}
 |\varphi \rangle \approx c_1 |z_1\rangle + c_2 |z_2\rangle \, .
     \label{aprowave}
\end{equation}
By substituting Eq.~(\ref{aprowave}) into Eq.~(\ref{ddconsv}), 
we obtain
\begin{equation}
       \frac{\rmd P(E)}{\rmd E} = |c_1|^2 |\langle E|z_1 \rangle |^2 +
   |c_2|^2 |\langle E|z_2 \rangle |^2 + 2 \, \text{Re} (c_1c_2^*
   \langle E|z_1 \rangle \langle E|z_2 \rangle ^*) \, .
      \label{ddconsvdd}
\end{equation}
It was shown in section~3 of Ref.~\cite{NPA15} that
\begin{equation}
       \langle E|z_1 \rangle = \frac{1}{z_1-E} \langle E|V|z_1\rangle \, ,
       \label{matridx}
\end{equation}
with an analogous result holding for $|z_2\rangle$. Substitution
of Eq.~(\ref{matridx}) into Eq.~(\ref{ddconsvdd}) yields
\begin{eqnarray}
       \frac{\rmd P(E)}{\rmd E} &=& |c_1|^2 
        \frac{1}{(E-E_1)^2+(\Gamma _1/2)^2}|\langle E|V|z_1 \rangle |^2+
   |c_2|^2 \frac{1}{(E-E_2)^2+(\Gamma _2/2)^2}|\langle E|V|z_2 \rangle |^2
        \nonumber \\
   && + 2 \, \text{Re} \left( c_1c_2^* \frac{1}{z_1-E}\frac{1}{z_2^*-E}
        \langle E|V|z_1 \rangle \langle E|V|z_2 \rangle ^* \right) \, .
      \label{ddconsvdd3}
\end{eqnarray}
This is the analog of Eq.~(\ref{crossecinter}). Both 
Eqs.~(\ref{crossecinter}) and~(\ref{ddconsvdd3}) contain the
contribution of each resonance plus an interference term. However,
in Eq.~(\ref{crossecinter}) the contribution from each resonance to 
the cross section is fixed and determined by the Breit-Wigner
distributions and the residues, whereas in Eq.~(\ref{ddconsvdd3}) the
contribution from each resonance can change depending on its weight ($c_1$ or
$c_2$) in the resonant expansion~(\ref{aprowave}).

It should be noted that the standard Golden Rule is 
unable to account for the interference of two resonances. The
reason is that the interference between two delta functions should be zero,
since there is no overlap between them. Thus, the standard Golden Rule
does not produce an interference term, and it is inadequate to describe
experiments such as, for example, that in Ref.~\cite{BESIII2}.

In the literature, one can find studies of the interference of two
resonances (see for example Refs.~\cite{PESKIN,ROTTER1,ROTTER2,SASADA}). The
approach of the present paper is basically the same as that in 
Ref.~\cite{SASADA}: One uses a resonant expansion~\cite{TOLSTIKHIN1,05CJP,TOLSTIKHIN2,TOLSTIKHIN3,GASTON12,HATANO14,BROWN16} to expand a square integrable 
wave function in terms of resonant states, and then one keeps the contributions
of the resonant states that carry the most weight in the expansion. It
is important to understand, however, that resonant 
expansions~\cite{TOLSTIKHIN1,05CJP,TOLSTIKHIN2,TOLSTIKHIN3,GASTON12,HATANO14,BROWN16} cannot be used to further expand the decay 
width, Eq.~(\ref{inteGam6n}), the decay constant, Eq.~(\ref{inteGam6totalti}), 
and the decay energy spectrum, Eq.~(\ref{defP}), of a single resonance. The 
formulas in
Eqs.~(\ref{inteGam6totaln}), (\ref{inteGam6totalti}) and~(\ref{defP}) already
provide the contribution of each individual resonance, and therefore cannot
be expanded any further in terms of the rest of the resonances of the system.

\section{Fits of resonant peaks}
\setcounter{equation}{0}
\label{sec:crosspeaks}

Fits of resonant peaks in cross sections are notoriously ambiguous, and 
one needs to recourse to additional quantities such as the phase shift
or the time delay to ascertain the presence and the exact location of a 
resonance (see for example Ref.~\cite{LUNAACOSTA}). When a system has an 
isolated resonance, it is usually assumed~\cite{TAYLOR} that in the vicinity 
of the resonant energy the $S$-matrix can be approximated by 
\begin{equation}
       S(E) \approx \frac{E-\zr ^*}{E-\zr} \, .
      \label{unitarie}
\end{equation}
Often, this approximation is written in terms of the wave number,
\begin{equation}
       S(E) \approx \frac{k-\kr ^*}{k-\kr} \, .
       \label{unitarik}
\end{equation}
Mathematically, the approximations~(\ref{unitarie}) and~(\ref{unitarik}) 
can be justified by way of Blaschke products. Physically, such
approximations are more desirable than the one resulting from neglecting 
the background in the Laurent expansion of the $S$-matrix, because the 
approximate
$S$-matrix in Eqs.~(\ref{unitarie}) and~(\ref{unitarik}) is unitary, whereas
the $S$-matrix that results from neglecting the background
in the Laurent expansion is not unitary. 

Because they are unitary, one may wonder if the 
approximations~(\ref{unitarie}) and~(\ref{unitarik}) lead to better
fits of resonant peaks than the Laurent-expansion approximation. In order
to find out, let us substitute Eq.~(\ref{unitarie}) into 
Eq.~(\ref{crossec}). The result is
\begin{equation}
      \sigma (E)\approx \sigma_{\text{\tiny e-unitarized}}(E)= \frac{\pi}{k^2} 
\frac{|\gr|^2}{(E-\er)^2+(\gr/2)^2} \, .
        \label{arppcue}
\end{equation}
Similarly, substitution of Eq.~(\ref{unitarik}) into Eq.~(\ref{crossec})
yields
\begin{equation}
      \sigma (E)\approx \sigma_{\text{\tiny k-unitarized}}(E)= \frac{\pi}{k^2} 
\frac{|2\br|^2}{(k-\ar)^2+\br ^2} \, .
        \label{arppcuk}
\end{equation}
Comparison of Eqs.~(\ref{arppc}), (\ref{arppcue}), and~(\ref{arppcuk})
shows that $\sigma_{\text{\tiny Laurent}}(E)$ and $\sigma_{\text{\tiny e-unitarized}}(E)$
are identical, except for an overall factor, whereas 
$\sigma_{\text{\tiny k-unitarized}}(E)$ is a Breit-Wigner distribution in the
wave number~\cite{BWM}, rather than in the energy, domain.

Figure~\ref{fig:peaks} shows the three approximations together with
the exact cross section in the vicinity of the third resonant energy for
$\lambda =100$. As can be seen in Fig.~\ref{fig:peaks}, the approximations
$\sigma_{\text{\tiny e-unitarized}}(E)$ and $\sigma_{\text{\tiny k-unitarized}}(E)$ are 
almost indistinguishable. In fact, their ratio is given by
\begin{equation}
     \frac{\sigma_{\text{\tiny e-unitarized}}(E)}{\sigma_{\text{\tiny k-unitarized}}(E)}
        = \frac{4\ar ^2}{(k+\ar )^2+\br ^2} \, ,
\end{equation}
which is very close to unity when $k$ is close to $\ar$ and when $\br$ is 
small, as is the case for most resonances of the delta-shell potential when
$|\lambda|$ is sufficiently large. One can also see in Fig.~\ref{fig:peaks} 
that the Laurent approximation $\sigma_{\text{\tiny Laurent}}(E)$ is very similar to
$\sigma_{\text{\tiny e-unitarized}}(E)$ and $\sigma_{\text{\tiny k-unitarized}}(E)$. Thus,
even though the corresponding approximation of the $S$-matrix is not 
unitary, $\sigma_{\text{\tiny Laurent}}(E)$ provides as good of a fit as
$\sigma_{\text{\tiny e-unitarized}}(E)$ and $\sigma_{\text{\tiny k-unitarized}}(E)$.

\section{Conclusions}
\setcounter{equation}{0}
\label{sec:conclusions}

The resonant (Gamow) states are the natural wave functions of 
resonances. Because they diverge exponentially at 
infinity, it is not easy to extract information from them. We have used
the delta-shell potential to show how one can calculate explicitly
the decay width $\overline{\Gamma}$ and the decay constant $\Gamma$ of a 
resonant state without having to worry about its exponential blowup
at infinity. We have 
also shown that the Golden Rule of a sharp resonant state yields a decay 
width that is approximately the same as the pole 
width, even though the exact decay width is very different from the pole width.
Overall, the results of the present paper constitute a numerical validation 
of the formalism of Ref.~\cite{NPA15}.

It should be noted that the decay width $\overline{\Gamma}$ and the decay
constant $\Gamma$ are not replacements of the pole width $\gr$, since 
$\gr$ determines the lifetime of the resonance. Rather, $\overline{\Gamma}$ and
$\Gamma$ provide another way to quantify the strength of the interaction between
the resonance and the continuum.

We have seen that for sharp resonances, the decay energy spectrum 
$\frac{\rmd P(E)}{\rmd E}$ has a shape that is similar to, but not the 
same as, the Breit-Wigner distribution, whereas 
for resonances that are not very sharp, the lineshape differs significantly
from the Breit-Wigner distribution, due to the effect of the
matrix element of the interaction. In particular, fits using the lineshape
$\frac{\rmd P(E)}{\rmd E}$ would yield different resonant energies and
pole widths than fits using only a Breit-Wigner distribution. We have also
pointed out some (vague) similarities between the decay energy spectrum of the
resonances of the delta-shell potential and some experimental decay energy
spectra.

In normalizing the distribution $\frac{\rmd P(E)}{\rmd E}$, we have found
that if we normalize a resonant state as 
$\frac{1}{\sqrt{\Gamma}}|\zr \rangle$, then 
$\frac{1}{\sqrt{\Gamma}}|\zr \rangle$ yields a normalized decay
energy spectrum. However, this normalization is not a replacement
of Zeldovich's normalization, since Zeldovich's normalization is more
appropriate for resonant expansions.

We have seen that
there is a clear analogy between the $S$-matrix description of a resonance
and the formalism of Ref.~\cite{NPA15}. Much like the pole of the $S$-matrix 
extracts the contribution of a resonance to the cross section through a 
Laurent expansion, the resonant state extracts the contribution of a
resonance to the decay energy spectrum through the expression 
for $\frac{\rmd P(E)}{\rmd E}$. Much like Lorentzian
peaks in the cross section are produced by intermediate, unstable particles,
the quasi-Lorentzian peaks in decay energy spectra are produced by
decaying, unstable particles. 

There are however some differences between cross sections and decay
energy spectra. For example, in the cross section, the contribution of 
each resonance is fixed, and it is 
determined by the residue of the $S$-matrix at the pole. In the decay energy 
spectrum the strength of each resonance's
contribution can change depending on the weight of that particular
resonant state in the resonant expansion.

We have interpreted the background term of resonant expansions as the
impossibility that a given experimental procedure used to create a 
resonant state is always
successful. That is, not even in principle a given
experimental procedure will yield a resonant state in all the trials. During
some trials, a non-resonant state will be created, and such
non-resonant state will be described by the background. However, in those
trials where we do create a resonant state, the wave function of such
state is a Gamow state, and the corresponding decay widths,
decay constants, and decay energy spectra are obtained in the
way presented in this paper.

We have argued that the standard Golden Rule is not appropriate to describe
the interference of two resonances, and we have used the decay energy
spectrum of two resonant states to describe such interference. The resulting
energy spectrum is very similar to the cross section obtained
by way of a Mittag-Leffler expansion.

We have also seen that three common approximations of the $S$-matrix
lead to similar fits of resonant peaks in the cross section.

Although strictly speaking our results are restricted to the resonances
of the delta-shell potential, it is likely that their main features hold true
for a large class of potentials that includes those of compact 
support. However, it would be interesting to see how our results apply
to more realistic 
systems~\cite{MICHEL7,KAMANO15,MIYAHARA16,KAMANOPRC16,KAMANOPRC16b,FLORES}.

\section*{Acknowledgments}
The author would like to thank Enriqueta Hernandez, Martin 
Tchernookov, Cengiz Sen, and George Irwin for enlightening discussions.

\appendix
\section{Appendix A: The Lambert $W$ function}
\setcounter{equation}{0}
\label{app:tlf}

The resonant wave numbers of a delta-shell potential can be expressed 
in terms of the Lambert $W$ 
function~\cite{LAMBERT1,LAMBERT2,LAMBERT3,LAMBERT4,LAMBERT5,LAMBERT6},
see Eq.~(\ref{complexsol}). In this Appendix, we will briefly
review the main properties of $W$, and we will show that 
Eq.~(\ref{complexsol}) provides the solutions to 
Eq.~(\ref{complexroots}).

The natural logarithm is the inverse function of the exponential function
$f(x)=\rme ^x$. That is,
\begin{equation}
       y=\rme ^x \quad \Longleftrightarrow \quad x=\ln (y) \, . 
       \label{defanlo}
\end{equation}
Similarly, the Lambert function is the inverse function of 
$f(x)=x\rme ^x$. That is,
\begin{equation}
       y=x\rme ^x \quad \Longleftrightarrow \quad x=W(y) \, . 
       \label{deflf}
\end{equation}

In order to obtain the solutions of Eq.~(\ref{complexroots}), let us 
define a new variable $t$ as~\cite{LAMBERT4}
\begin{equation}
     t=-2\rmi a k + \lambda \, .
    \label{definoft}
\end{equation}
Using this new variable, Eq.~(\ref{complexroots}) can be written as
\begin{equation}
    t\rme ^t = \lambda  \rme ^{\lambda } \, .
\end{equation} 
Hence, by the definition~(\ref{deflf}) of the Lambert function, we have
that
\begin{equation}
      t=W(\lambda  \rme ^{\lambda}) \, .
       \label{sefd}
\end{equation}
Substitution of Eq.~(\ref{definoft}) into Eq.~(\ref{sefd}) yields
\begin{equation}
        k=\frac{1}{2 \rmi a} \left[ \lambda  -  
         W(\lambda  \rme ^{\lambda })\right] \, ,
        \label{genrasold}
\end{equation}
which is just Eq.~(\ref{complexsol}). The branches of the Lambert function 
generate, through Eq.~(\ref{genrasold}), the resonant, anti-resonant, 
bound, and virtual poles of the delta-shell potential. It should be noted that
Eq.~(\ref{genrasold}) also produces the poles of the complex delta 
potential~\cite{GASTONPRA14}.

\section{Appendix B: Tables}
\setcounter{equation}{0}
\label{app:tables}

In this appendix, we list the resonant quantities of the first
eight resonances of the delta-shell potential for 
$\lambda = \pm 0.5, \pm 10$, and $\pm 100$. The wave number
$\kr$ is in units of $1/a$, whereas the resonant energies
$\zr$, the pole width $\gr$, and the decay widths 
$\overline{\Gamma}$ and $\overline{\Gamma}_{\rm sharp}$ are in
units of $\hbar ^2/2ma^2$. The decay constants
$\Gamma$ and $\Gamma_{\rm sharp}$ are dimensionless.

\vskip-0.3cm

\begin{table}[H]
\centering
\vspace{6pt}
\begin{tabular}{|l| c| c| c| c | c | c | c|} % centered columns (4 columns)
\hline\hline
      & $\kr$ & $\zr$ & $\gr$ & $\overline{\Gamma}$ & $\Gamma$ & 
$\overline{\Gamma}_{\rm sharp}$
& $\Gamma_{\rm sharp}$  \\

\hline

1st & $3.1105 -0.000956 \, \rmi$ & $9.6754 -0.00595 \, \rmi$ & 
0.0119 & 0.0237 & 1.9924 & 0.0119 & 0.9972    \\

2nd & $6.2213 -0.003803 \, \rmi$ & $38.704 -0.04732 \, \rmi$ & 
0.0946 & 0.1864 & 1.9700 & 0.0936 & 0.9888  \\

3rd & $9.3325  -0.008479 \, \rmi$ & $87.096 -0.15827 \, \rmi$ & 
0.3165 & 0.6121 & 1.9337 & 0.3087 & 0.9752  \\

4th & $12.444 - 0.014885 \, \rmi$ & $154.86 -0.37048 \, \rmi$ & 
0.7410 & 1.3969 & 1.8852 & 0.7090 & 0.9569  \\

5th & $15.557 -0.022893 \, \rmi$ & $242.02 -0.71228 \, \rmi$ & 
1.4246 & 2.6018 & 1.8264 & 1.3313 & 0.9345 \\

6th & $18.671 -0.032350 \, \rmi$ & $348.59 -1.20800 \, \rmi$ & 
2.4159 & 4.2509 & 1.7595 & 2.1956 & 0.9088  \\

7th & $21.786 -0.043093 \, \rmi$ & $474.61 -1.87762 \, \rmi$ & 
3.7552 & 6.3338 & 1.6867 & 3.3060 & 0.8804  \\

8th & $24.902 -0.054952 \, \rmi$ & $620.08 -2.73677 \, \rmi$ & 
5.4735 & 8.8126 & 1.6100 & 4.6530 & 0.8501  \\

\hline\hline 
\end{tabular}
\caption{Resonant quantities for $\lambda =100$. }
\label{table:100}
\end{table}

\vskip-0.3cm

\begin{table}[!h]
\centering
\vspace{6pt}
\begin{tabular}{|l| c| c| c| c | c | c | c|} % centered columns (4 columns)
\hline\hline
      & $\kr$ & $\zr$ & $\gr$ & $\overline{\Gamma}$ & $\Gamma$ & 
$\overline{\Gamma}_{\rm sharp}$
& $\Gamma_{\rm sharp}$  \\
\hline
1st &  $2.8776 -0.06651 \, \rmi$  & $8.2760 -0.3828\, \rmi$ & 
$0.7656$  & $1.1887$ & $1.5527$ & $0.6416$ & $0.8381$ \\

2nd & $5.8413 -0.20648\, \rmi$ & $34.079 -2.4123\, \rmi$ & 
$4.8245$ & $4.4336$ & $0.9190$ & $2.7565$ & $0.5713$ \\

3rd & $8.8807\, -0.34784\, \rmi$ & $78.745-6.1781\, \rmi$ & 
$12.356$ & $6.7976$ & $0.5501$ & $4.7517$ & $0.3846$  \\

4th & $11.962 -0.46966\, \rmi$ & $142.87\ -11.236\, \rmi$ & 
22.472 & 8.0742 & 0.3593 & 6.1270 & 0.2726  \\

5th & $15.066 -0.57220\, \rmi$ & $226.64 -17.241\, \rmi$ & 
34.482 & 8.7614 & 0.2541 & 7.0385 & 0.2041  \\

6th & $18.181 -0.65934 \, \rmi$ & $330.13 -23.975\, \rmi$ & 
47.951 & 9.1543 & 0.1909 & 7.6599 & 0.1597  \\

7h & $21.305 -0.73457\, \rmi$ & $453.34 -31.299 \rmi$ & 
62.599 & 9.3939 & 0.1501 & 8.1005 & 0.1294  \\

8th & $24.432 -0.80052\, \rmi$ & $596.30 -39.117 \rmi$ & 
78.234 & 9.5484 & 0.1220 & 8.4245 & 0.1077  \\

\hline\hline 
\end{tabular}
\caption{Resonant quantities for $\lambda =10$.}
\label{table:10}
\end{table}

\vskip-0.3cm

%\vspace*{5cm}

\begin{table}[ht!]
\centering
\vspace{6pt}
\begin{tabular}{|l| c| c| c| c | c | c | c|c|c|} % centered columns (4 columns)
\hline\hline
      & $\kr$ & $\zr$ & $\gr$ & $\overline{\Gamma}$ & $\Gamma$ & 
$\overline{\Gamma}_{\rm sharp}$
& $\Gamma_{\rm sharp}$  \\
\hline

1st & $2.1659 -1.1167 \, \rmi$ & $3.4440 -4.8372 \, \rmi$ & 
9.6745 & 0.53790 & 0.05560 & 1.34145 & 0.13866  \\

2nd & $5.3794 -1.5486 \, \rmi$ & $26.539 -16.661 \, \rmi$ & 
33.322 & 0.50608 & 0.01519 & 0.91622 & 0.02750  \\

3rd & $8.5512 -1.7740 \, \rmi$ & $69.976 -30.340 \, \rmi$ & 
60.679 & 0.50188 & 0.00827 & 0.80531 & 0.01327  \\

4th & $11.710 -1.9284 \, \rmi$ & $133.40 -45.163 \, \rmi$ & 
90.327 & 0.50069 & 0.00554 & 0.74903 & 0.00829  \\

5th & $14.862 -2.0462 \, \rmi$ & $216.71 -60.823 \, \rmi$ & 
121.65 & 0.50024 & 0.00411 & 0.71355 & 0.00587  \\

5th & $18.012 -2.1414 \, \rmi$ & $319.84 -77.141 \, \rmi$ & 
154.28 & 0.50005 & 0.00324 & 0.68860 & 0.00446  \\

7th & $21.160 -2.2213 \, \rmi$ & $442.78 -94.004 \, \rmi$ & 
188.01 & 0.49995 & 0.00266 & 0.66985 & 0.00356 \\

8th & $24.306 -2.2903 \, \rmi$ & $585.51 -111.33 \, \rmi$ & 
222.66 & 0.49991 & 0.00225 & 0.65513 & 0.00294  \\

\hline\hline 
\end{tabular}
\caption{Resonant quantities for $\lambda =0.5$.}
\label{table:0.5}
\end{table}

\newpage

\begin{table}[H]
\centering
\vspace{6pt}
\begin{tabular}{|l| c| c| c| c | c | c |c |} % centered columns (4 columns)
\hline\hline
      & $\kr$ & $\zr$ & $\gr$ & $\overline{\Gamma}$ & $\Gamma$ & 
$\overline{\Gamma}_{\rm sharp}$
& $\Gamma_{\rm sharp}$ \\
\hline

virt. &  $-0.6282 \, \rmi$ & 
$-0.3947$ & 0 & 0 & 0.18817 & &   \\

1st & $3.7188 -1.3945 \, \rmi$ & $11.885 -10.372 \, \rmi$ & 
20.744 & 0.45592 & 0.02198 & 0.10662 & 0.00514  \\

2nd & $6.9328 -1.6799 \, \rmi$ & $45.242 -23.294 \, \rmi$ & 
46.587 & 0.48116 & 0.01033 & 0.19640 & 0.00422 \\

3rd &  $10.107 -1.8604 \, \rmi$ & $98.695 -37.608 \, \rmi$ & 
75.216 & 0.48917 & 0.00650 & 0.24735 & 0.00329  \\

4th & $13.268 -1.9929\, \rmi$ & $172.07 -52.883 \, \rmi$ & 
105.77 & 0.49284 & 0.00466 & 0.28122 & 0.00266  \\

5th & $16.422 -2.0975 \, \rmi$ & $265.30 -68.893 \, \rmi$ & 
137.79 & 0.49486 & 0.00359 & 0.3057 & 0.00222  \\

6th & $19.573 -2.1841 \, \rmi$ & $378.34 -85.498 \, \rmi$ & 
171.00 & 0.49610 & 0.00290 & 0.32448 & 0.00190 \\

7th & $22.722 -2.2578 \, \rmi$ & $511.17 -102.60\, \rmi$ & 
205.21 & 0.49692 & 0.00242 & 0.33937 & 0.00165  \\

8th & $25.869 -2.3221 \, \rmi$ & $663.79 -120.14 \, \rmi$ & 
240.28 & 0.49750 & 0.00207 & 0.35154 & 0.00146  \\

\hline\hline 
\end{tabular}
\caption{Resonant quantities for $\lambda =-0.5$.} 
\label{table:-0.5}
\end{table}

\begin{table}[H]
\centering
\vspace{6pt}
\begin{tabular}{|l| c| c| c| c | c | c | c|} % centered columns (4 columns)
\hline\hline
      & $\kr$ & $\zr$ & $\gr$ & $\overline{\Gamma}$ & $\Gamma$ & 
$\overline{\Gamma}_{\rm sharp}$
& $\Gamma_{\rm sharp}$  \\
\hline

bound & $4.9998 \, \rmi$ &  $-24.998$ & 
0 & 0 & 1 &  &     \\

1st & $3.4380 -0.1038 \, \rmi$ & $11.809 -0.7138 \, \rmi$ & 
1.4276 & 1.8693 & 1.3094 & 1.0118 & 0.7087  \\

2nd &  $6.7367 -0.2684 \, \rmi$ & $45.312 -3.6159 \, \rmi$ & 
7.2318 & 5.0728 & 0.7015 & 3.0729 & 0.4249 \\

3rd & $9.9614 -0.4092 \, \rmi$ & $99.063 -8.1533 \, \rmi$ & 
16.307 & 6.9389 & 0.4255 & 4.5690 & 0.2802  \\

4th & $13.153 -0.5242 \, \rmi$ & $172.73 -13.790 \, \rmi$ & 
27.580 & 7.9535 & 0.2884 & 5.5583 & 0.2015  \\

5th &  $16.328 -0.6197 \, \rmi$ & $266.21 -20.236 \, \rmi$ & 
40.473 & 8.5436 & 0.2111 & 6.2386 & 0.1541  \\

6th & $19.493 -0.7008 \, \rmi$ & $379.47 -27.322 \, \rmi$ & 
54.643 & 8.9122 & 0.1631 & 6.7301 & 0.1232  \\

7th & $22.652 -0.7711 \, \rmi$ & $512.51 -34.935 \, \rmi$ & 
69.869 & 9.1567 & 0.1311 & 7.1005 & 0.1016  \\

8th & $25.807 -0.8331 \, \rmi$ & $665.31 -42.998 \, \rmi$ & 
85.995 & 9.3268 & 0.1085 & 7.3895 & 0.0859  \\

\hline\hline 
\end{tabular}
\caption{Resonant quantities for $\lambda =-10$.} 
\label{table:-10}
\end{table}

\begin{table}[H]
\centering
\vspace{6pt}
\begin{tabular}{|l| c| c| c| c | c | c | c|c|c|} % centered columns (4 columns)
\hline\hline
      & $\kr$ & $\zr$ & $\gr$ & $\overline{\Gamma}$ & $\Gamma$ & 
$\overline{\Gamma}_{\rm sharp}$
& $\Gamma_{\rm sharp}$   \\
\hline

bound &  $50 \, \rmi$ &  $-2500$ & 
0 & 0 & 1 &  &    \\

1st & $3.1733 -0.00102 \, \rmi$ & $10.070 -0.0064 \, \rmi$ & 
0.0129 & 0.0257 & 1.9919 & 0.0128 & 0.9969  \\

2nd & $6.3463 -0.00404 \, \rmi$ & $40.276 -0.0512 \, \rmi$ & 
0.1024 & 0.2016 & 1.9678 & 0.1012 & 0.9878  \\

3rd & $9.5188 -0.00899 \, \rmi$ & $90.608 -0.1711 \, \rmi$ & 
0.3422 & 0.6601 & 1.9291 & 0.3330 & 0.9731  \\

4th & $12.691 -0.01576 \, \rmi$ & $161.05 -0.3999 \, \rmi$ & 
0.7998 & 1.5017 & 1.8776 & 0.7624 & 0.9533  \\

5th & $15.862 -0.02419 \, \rmi$ & $251.59 -0.7674 \, \rmi$ & 
1.5349 & 2.7864 & 1.8154 & 1.4262 & 0.9292  \\

6th & $19.031 -0.03412 \, \rmi$ & $362.19 -1.2988 \, \rmi$ & 
2.5976 & 4.5330 & 1.7451 & 2.3423 & 0.9017  \\

7th & $22.200 -0.04536 \, \rmi$ & $492.83 -2.0140 \, \rmi$ & 
4.0281 & 6.7233 & 1.6691 & 3.5110 & 0.8716  \\

8th & $25.367 -0.05772 \, \rmi$ & $643.50 -2.9284 \, \rmi$ & 
5.8568 & 9.3104 & 1.5897 & 4.9182 & 0.8397  \\

\hline\hline 
\end{tabular}
\caption{Resonant quantities for $\lambda =-100$.}
\label{table:-100}
\end{table}

\newpage

\section{Appendix C: Plots}
\setcounter{equation}{0}
\label{app:plots}

\begin{figure}[h!]
\begin{center}
              \epsfxsize=11cm
              \epsffile{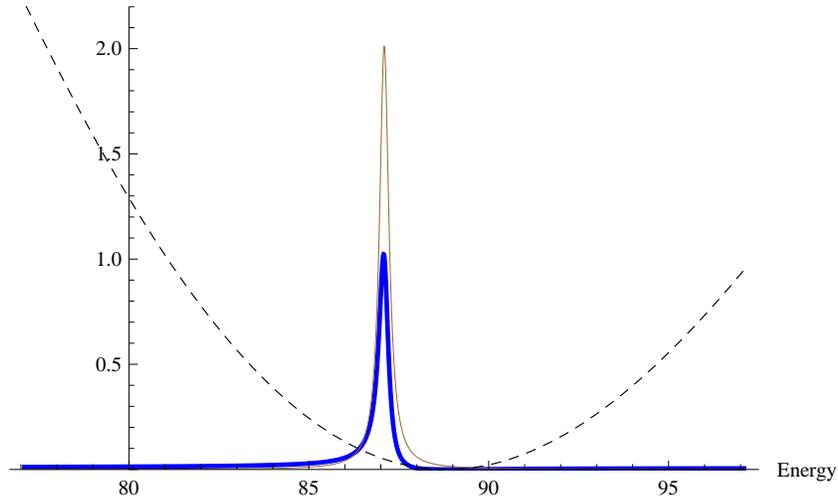}
\end{center}                
\caption{Plot of the decay energy spectrum (blue, thick, solid line), 
the normalized Breit-Wigner distribution (brown, thin, solid line), and the 
matrix element of the interaction (dashed, black line)
for the third resonance when $\lambda =100$.}
\label{fig:thirdthree100}
\end{figure}

\begin{figure}[h!]
\begin{center}
              \epsfxsize=11cm
              \epsffile{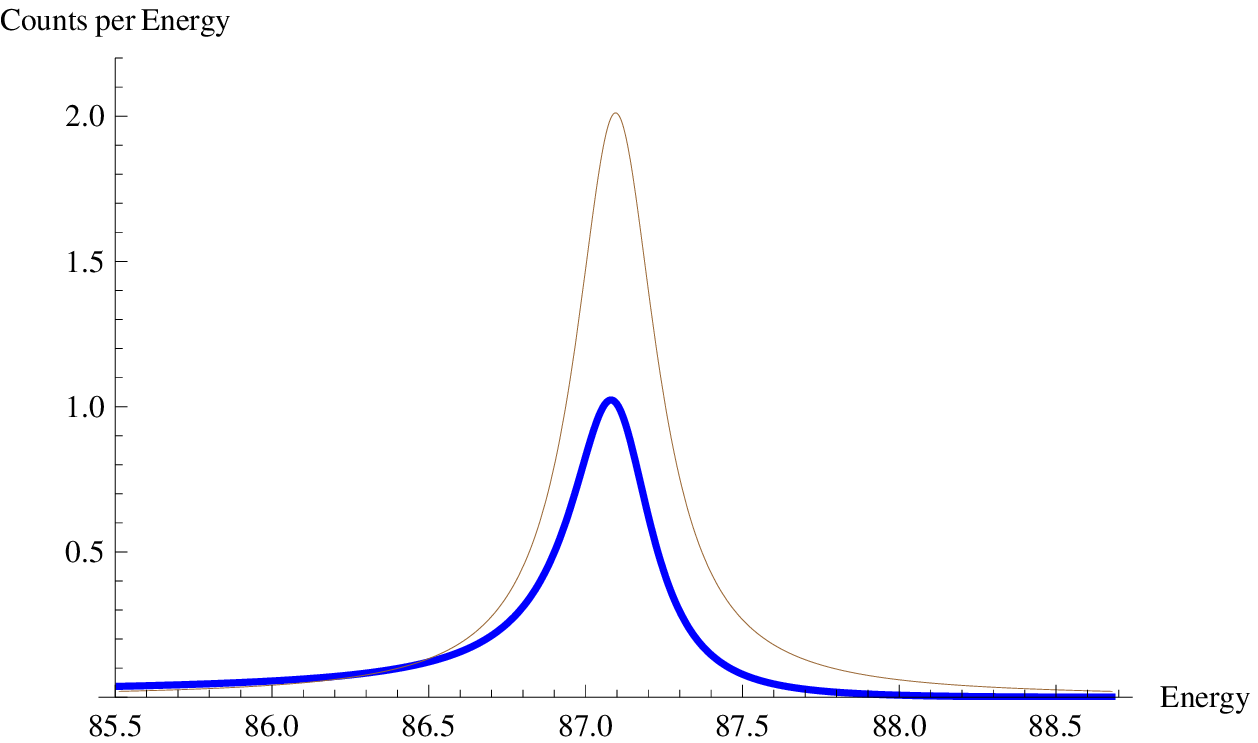}
\end{center}                
\caption{Close-up of the decay energy spectrum (blue, thick, solid line), 
and the normalized Breit-Wigner distribution (brown, thin, solid line) 
for the third resonance when $\lambda =100$.}
\label{fig:thirdtwo100}
\end{figure}

\begin{figure}[h!]
\begin{center}
              \epsfxsize=11cm
              \epsffile{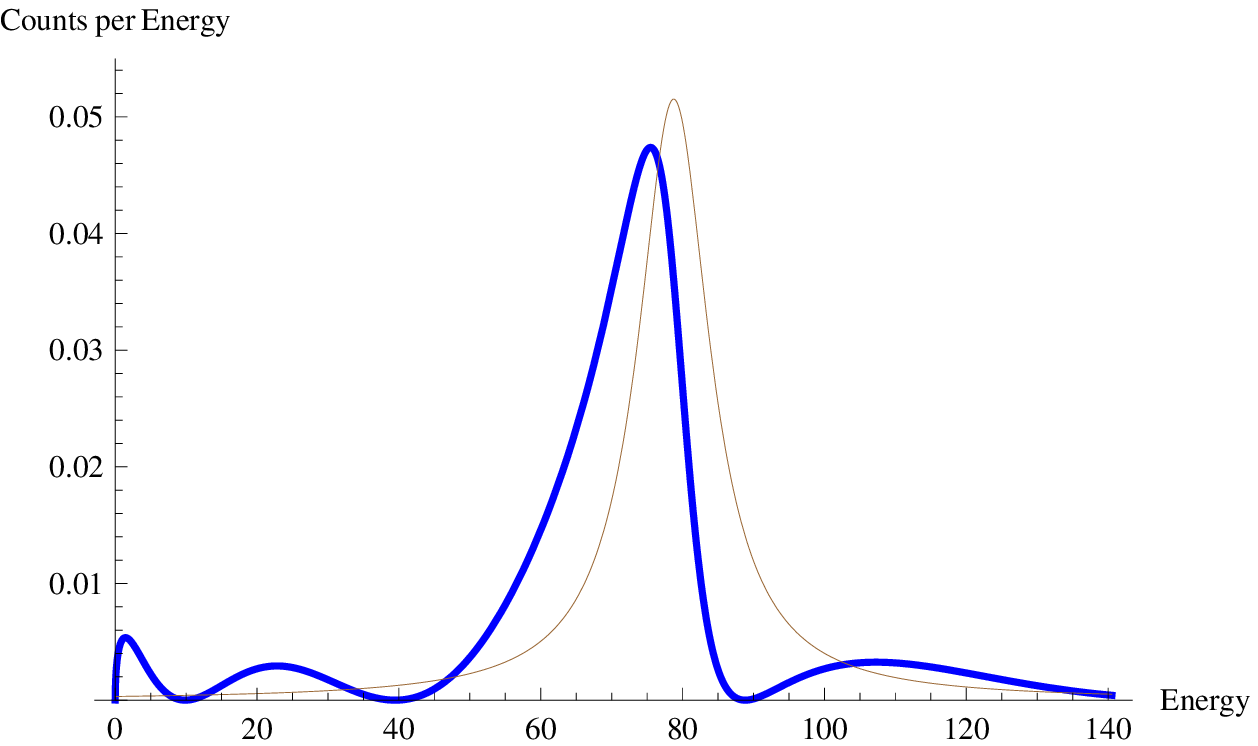}
\end{center}                
\caption{Plot of the decay energy spectrum (blue, thick, solid line), and 
the normalized Breit-Wigner distribution (brown, thin, solid line) for 
the third resonance when $\lambda =10$.}
\label{fig:thirdtwo10}
\end{figure}

\quad

\vskip2cm

\begin{figure}[H]
\begin{center}
              \epsfxsize=11cm
              \epsffile{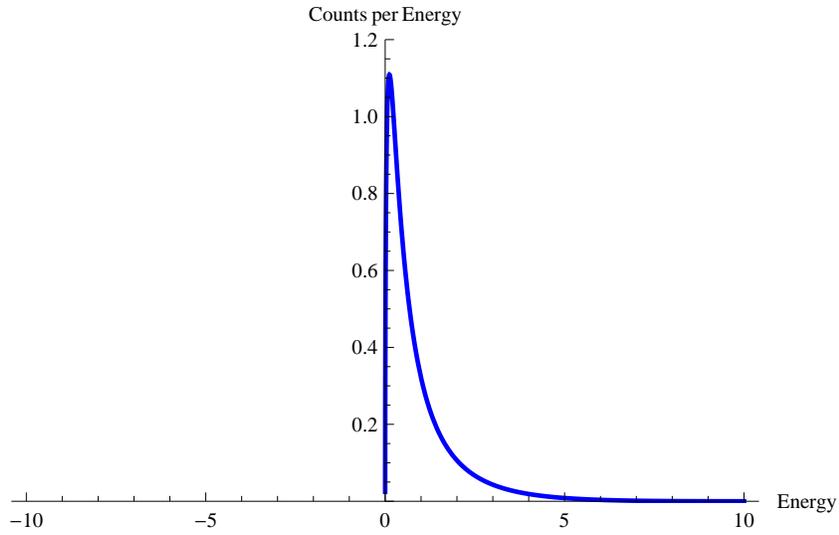}
\end{center}                
\caption{Plot of the decay energy spectrum of the virtual state 
when $\lambda =-0.5$.}
\label{fig:esdvirtual}
          \end{figure}

\begin{figure}[H]
\begin{center}
              \epsfxsize=11cm
              \epsffile{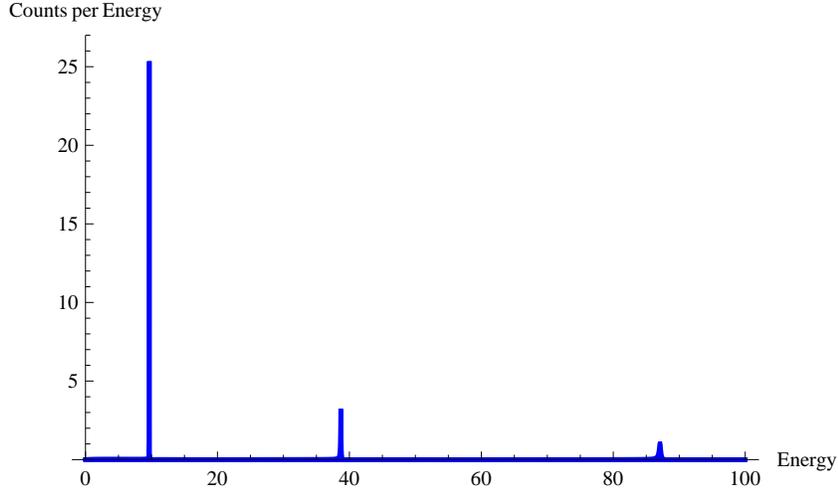}
\end{center}                
\caption{Plot of the decay energy spectra of the first three
resonances when $\lambda =100$.}
\label{fig:spikes}
\end{figure}

\begin{figure}[H]
\begin{center}
              \epsfxsize=12cm
              \epsffile{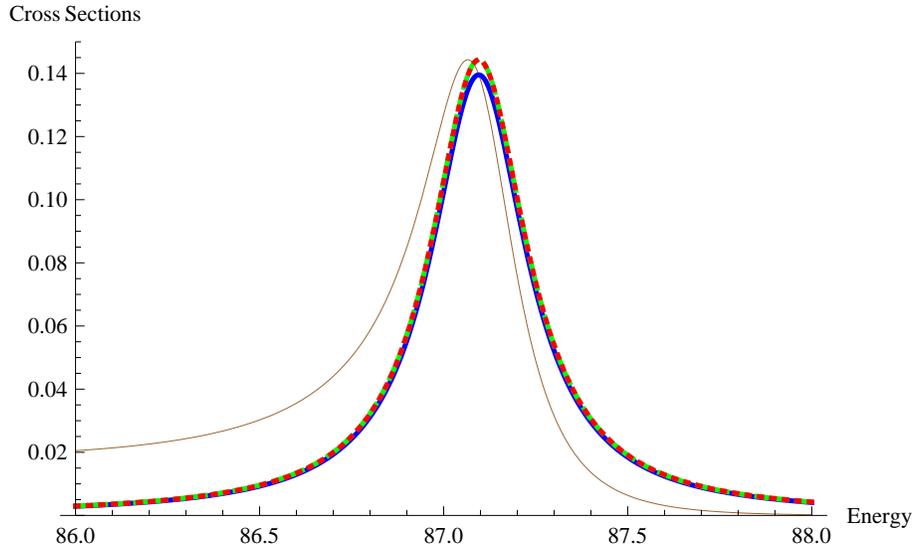}
\end{center}                
\caption{Plots of the exact cross section $\sigma (E)$ (thin, brown line), 
the Laurent approximation of the cross section $\sigma_{\text{\tiny Laurent}}(E)$
(thick, blue line), the e-unitarized approximation of the cross section
$\sigma_{\text{\tiny e-unitarized}}(E)$ (dashed, green line), and the k-unitarized
approximation of the cross section $\sigma_{\text{\tiny k-unitarized}}(E)$
(dotted, red line) in the vicinity of the third resonant energy for
$\lambda =100$. For this resonance, $\sigma_{\text{\tiny e-unitarized}}(E)$
and $\sigma_{\text{\tiny k-unitarized}}(E)$ are virtually indistinguishable.}
\label{fig:peaks}
\end{figure}


\begin{thebibliography}{99}



\bibitem{TAYLOR} J.R.~Taylor, {\it Scattering Theory}, John Wiley \& Sons
(1972). 

\bibitem{SIRLIN} A.~Sirlin, Phys.~Rev.~Lett.~{\bf 67}, 2127 (1991).

\bibitem{WILLENBROCK} S.~Willenbrock and G.~Valencia, 
Phys.~Lett.~B~{\bf 259}, 373 (1991).

\bibitem{STUART} R.G.~Stuart, Phys.~Lett.~B~{\bf 262}, 113 (1991).

\bibitem{LEIKE} A.~Leike, T.~Riemann, and J.~Rose, Phys.~Lett.~B~{\bf 273}, 
513 (1991).

\bibitem{BERNICHA} A.~Bernicha, G.~Lopez Castro, and J.~Pestieau, 
Phys.~Rev.~D~{\bf 50}, 4454 (1994); 
Nucl.~Phys.~A~{\bf 597}, 623 (1996).

\bibitem{CASO} Particle Data Group, S.~Eidelman {\it et al.}, Phys.~Lett.~B
{\bf 592}, 1 (2004).

\bibitem{BS} A.R.~Bohm, Y.~Sato, Phys.~Rev.~D~{\bf 71},
085018 (2005).

\bibitem{ELANDER1} S.A.~Rakityansky, N.~Elander, 
Int.~J.~Quant.~Chemistry~{\bf 106} 1105 (2006); {\sf arXiv:1110.4990}.

\bibitem{GEGELIA} D.~Djukanovic, J.~Gegelia, S.~Scherer, 
Phys.~Rev.~D~{\bf 76}, 037501 (2007); {\sf arXiv:0707.2030}

\bibitem{ELANDER4} S.A.~Rakityansky, S.A.~Sofianos, N.~Elander, 
J.~Phys.~A: Math.~Theor.~{\bf 40}, 14857 (2007); {\sf arXiv:1110.4988}.

\bibitem{ELANDER3} K.~Shilyaeva, E.~Yarevsky, N.~Elander, 
J.~Phys.~B: Atom.~Mol.~Opt.~Phys.~{\bf 42}, 044011 (2009).

\bibitem{CECI2} M.~Had\v{z}imehmedovi\'c, S.~Ceci, 
A.~\v{S}varc, H.~Osmanovi\'c, J.~Stahov, Phys.~Rev.~C~{\bf 84}, 035204 (2011);
{\sf arXiv:1103.2653}.

\bibitem{CECI1} S.~Ceci, M.~Korolija, B.~Zauner, 
Phys.~Rev.~Lett.~{\bf 111}, 112004 (2013); {\sf arXiv:1302.3491}.

\bibitem{TIATOR13} R.L.~Workman, L.~Tiator, A.~Sarantsev
Phys.Rev.~C~{\bf 87}, 068201 (2013); {\sf arXiv:1304.4029}.

\bibitem{PDG} K.A.~Olive et al.~(Particle Data Group), 
Chin.~Phys.~C~{\bf 38}, 090001 (2014).

\bibitem{TIATOR2} A.~\v{S}varc, M.~Had\v{z}imehmedovi\'c, 
H.~Osmanovi\'c, J.~Stahov, L.~Tiator, R.L.~Workman,
Phys.~Rev.~C~{\bf 89}, 065208 (2014); {\sf arXiv:1404.1544}.

\bibitem{Vaandrager} P.~Vaandrager, S.A.~Rakityansky, 	
Int.~J.~Mod.~Phys.~E, {\bf 25}, 1650014 (2016); {\sf arXiv:1603.01718}.

\bibitem{TIATOR3} L.~Tiator, M.~D\"oring, R.L.~Workman, 
M.~Had\v{z}imehmedovi\'c, H.~Osmanovi\'c, R.~Omerovi\'c, J.~Stahov, 
A.~\v{S}varc, Phys.~Rev.~C~{\bf 94}, 065204 (2016); {\sf arXiv:1606.00371}.

\bibitem{GAMOW} G.~Gamow, Z.~Phys.~{\bf 51}, 204 (1928). 

\bibitem{SIEGERT} A.F.J.~Siegert, Phys.~Rev.~{\bf 56},
750 (1939). 

\bibitem{ZELDOVICH} Ya.B.~Zeldovich, Sov.~Phys.~JETP~{\bf 12}, 542 (1961).

\bibitem{BERGGREN} T.~Berggren, Nucl.~Phys.~A~{\bf 109}, 265 (1968).

\bibitem{PESKIN} U.~Peskin, H.~Reisler, W.H.~Miller, J.~Chem.~Phys.~{\bf 101}, 
9672 (1994).

\bibitem{BOLLINI1} C.G.~Bollini, O.~Civitarese, A.L.~De Paoli, M.C.~Rocca,
Phys.~Lett.~B{\bf 382}, 205 (1996).

\bibitem{TOLSTIKHIN1} O.I.~Tolstikhin, V.N.~Ostrovsky, H.~Nakamura,
Phys.~Rev.~A~{\bf 58}, 2077 (1998).

\bibitem{MONDRAGON00} E.~Hern\'andez, A.~J\'auregui, A.~Mondrag\'on, 
J.~Phys.~A: Math.~Gen.~{\bf 33}, 4507 (2000).

\bibitem{05CJP} R.~de la Madrid, G.~Garcia-Calderon, J.G.~Muga,
Czech.\ J.\ Phys.~{\bf 55}, 1141 (2005); {\sf quant-ph/0512242}.

\bibitem{TOLSTIKHIN2} O.I.~Tolstikhin, Phys.~Rev.~A~{\bf 73}, 062705 (2006).

\bibitem{TOLSTIKHIN3} O.I.~Tolstikhin, Phys.~Rev.~A~{\bf 77}, 032712 (2008).

\bibitem{VELAZQUEZ} J.M.~Velazquez-Arcos, C.A.~Vargas, 
J.L.~Fernandez-Chapou, A.L.~Salas-Brito, 
J.~Math.~Phys.~{\bf 49}, 103508 (2008).

\bibitem{NPA08} R.~de la Madrid, Nucl.~Phys.~A~{\bf 812}, 13 (2008);
{\sf arXiv:0810.0876}.

\bibitem{MICHEL7} N.~Michel, W.~Nazarewicz, M.~Ploszajczak, T.~Vertse, 
J.~Phys.~G: Nucl.~Part.~Phys.~{\bf 36}, 013101
(2009); {\sf arXiv:0810.2728}.

\bibitem{NIMROD} T.~Goldzak, I.~Gilary, N.~Moiseyev,
Phys.~Rev.~A~{\bf 82}, 052105 (2010).

\bibitem{GASTON10} G.~Garcia-Calderon, 
Advances in Quantum Chemistry~{\bf 60}, 407 (2010).

\bibitem{SASADA} K.~Sasada, N.~Hatano, G.~Ordonez, J.~Phys.~Soc.~Jpn.~{\bf 80},
104707 (2011); {\sf arXiv:0905.3953}.

\bibitem{GASTON11} G.~Garcia-Calderon, L.G.~Mendoza-Luna, 
Phys.~Rev.~A~{\bf 84}, 032106 (2011); {\sf arXiv:1104.4688}.

\bibitem{GASTON12} G.~Garcia-Calderon, A.~Mattar, J.~Villavicencio,
Phys.~Scr.~T{\bf 151}, 01476 (2012); {\sf arXiv:1205.0487}.

\bibitem{GASTON13} L.~Chaos-Cador, G.~Garcia-Calderon, Phys.~Rev.~A,
{\bf 87}, 042114 (2013).

\bibitem{FOSSEZ} K.~Fossez, N.~Michel, W.~Nazarewicz, M.~P{\l}oszajczak,
Phys.~Rev.~A, {\bf 87}, 042515 (2013); {\sf arXiv:1303.1928}.

\bibitem{ANDERSEN} C.K.~Andersen, K.~M{\o}lmer, Phys.~Rev.~A~{\bf 87}, 
052119 (2013); {\sf arXiv:1303.5644}.

\bibitem{JULVE14} J.~Julve, S.~Turrini, F.J.~de Urr{\'i}es,
Int.~J.~Theo.~Phys.~{\bf 53}, 971 (2014); {\sf arXiv:1302.0630}.

\bibitem{HATANO14} N.~Hatano, G.~Ordonez, J.~Math.~Phys.~{\bf 55}
122106 (2014); {\sf arXiv:1405.6683}.

\bibitem{GASTONPRA14} G.~Garcia-Calderon, L.~Chaos-Cador,
Phys.~Rev.~A~{\bf 90}, 032109 (2014); {\sf arXiv:1605.00999}

\bibitem{SARA} S.~Cruz y Cruz, O.~Rosas-Ortiz, 
Adv.~Math.~Phys.~281472 (2015); {\sf arXiv:1504.01008}.

\bibitem{LUNDMARK} R.~Lundmark, C.~Forss\'en, J.~Rotureau, 
Phys.~Rev.~A, {\bf 91}, 041601 (2015); {\sf arXiv:1412.7175}.

\bibitem{GENTILINI} S.~Gentilini, M.C.~Braidotti, G.~Marcucci, 
E.~DelRe, C.~Conti, Phys.~Rev.~A, {\bf 92}, 023801 (2015);
{\sf arXiv:1508.00692}.

\bibitem{NPA15} R.~de la Madrid, Nucl.~Phys.~A~{\bf 940}, 297 (2015);
{\sf arXiv:1505.07139}.

\bibitem{GASTON16} G.~Garcia-Calderon, R.~Romo, 
Phys.~Rev~A, {\bf 93}, 022118 (2016).

\bibitem{BROWN16} J.M.~Brown, P.~Jakobsen, A.~Bahl, J.V.~Moloney, M.~Kolesik, 
J.~Math.~Phys.~{\bf 57}, 032105 (2016).

\bibitem{PLASTINO1} A.~Plastino, M.C.~Rocca, 
Nucl.~Phys.~A, {\bf 948}, 19 (2016); {\sf arXiv:1511.04010}.

\bibitem{PLASTINO2} A.~Plastino, M.C.~Rocca, D.J.~Zamora
Nucl.~Phys.~A, {\bf 955}, 16 (2016); {\sf arXiv:1604.06910}.

\bibitem{CEVIK} D.~Cevik, M.~Gadella, S.~Kuru, J.~Negro,
Phys.~Lett.~A~{\bf 380}, 1600 (2016); {\sf arXiv:1601.05134}.

\bibitem{OLENDSKI} O.~Olendski, Ann.~Phys.~(Berlin), {\bf 529} 1600144 (2017);
{\sf arXiv:1611.00197}.

\bibitem{GARMON} S.~Garmon, G.~Ordonez, {\sf arXiv:1609.07718}.

\bibitem{GASTON17} G.~Garcia-Calderon, L.~Chaos-Cador,
Fortschritte der Physik (to be published); {\sf arXiv:1702.02247}.

\bibitem{DUNCAN} D.~Carlsmith, {\it Particle Physics}, Pearson (2012).

\bibitem{WINTER} R.G.~Winter, Phys.~Rev.~{\bf 123}, 1503 (1961).

\bibitem{GOTTFRIED} K.~Gottfried, {\it Quantum Mechanics}, 
W.A.~Benjamin (1966).

\bibitem{DICUS} D.A.~Dicus, W.W.~Repko, R.F.~Schwitters, T.M.~Tinsley,
Phys.~Rev.~A~{\bf 65}, 032116 (2002).

\bibitem{SCOTT} T.C.~Scott, J.F.~Babb, A.~Dalgarno, J.D.~Morgan~III,
J.~Chem.~Phys.~{\bf 99} 2841 (1993).

\bibitem{SANTINI1} U.G.~Aglietti, P.M.~Santini, Phys.~Rev.~A~{\bf 89}, 
022111 (2014); {\sf arXiv:1303.4977}.

\bibitem{SANTINI2} U.G.~Aglietti, P.M.~Santini, 
J.~Math.~Phys.~{\bf 56}, 062104 (2015); {\sf arXiv:1503.02532}.

\bibitem{SEGRE} E.~Segr\`e, {\it Nuclei and Particles: An Introduction 
to Nuclear and Subnuclear Physics,} Benjamin-Cummings Publishing Company, 
2nd edition (1977).


\bibitem{LAMBERT1} J.H.~Lambert, Acta Helveticae 
physico-mathematico-anatomico-botanico-medica, Band III, 128 (1758). 

\bibitem{LAMBERT2} L.~Euler, Acta Acad.~Scient.~Petropol.~{\bf 2}, 29 
(1783). 

\bibitem{LAMBERT3} R.M.~Corless, G.H.~Gonnet, D.E.G.~Hare,
D.J.~Jeffrey, D.E.~Knuth, 
Advances in Computational Mathematics~{\bf 5}, 329 (1996).

\bibitem{LAMBERT4} https://en.wikipedia.org/wiki/Lambert$_-$W$_-$function.

\bibitem{LAMBERT5} http://mathworld.wolfram.com/LambertW-Function.html

\bibitem{LAMBERT6} A.M.~Ishkhanyan, Phys.~Lett.~A~{\bf 380}, 640 (2016);
{\sf arXiv:1509.00846}.

\bibitem{NOTE} Since
the Lambert $W$ function has rarely been used in the physics literature,
we devote Appendix~\ref{app:tlf} to show that the solutions of
Eq.~(\ref{complexroots}) can indeed be written in terms of $W$.

\bibitem{ANDREASSEN1} A.~Andreassen, D.~Farhi, W.~Frost, M.D.~Schwartz,
Phys.~Rev.~Lett.~{\bf 117}, 231601 (2016); {\sf arXiv:1602.01102}.

\bibitem{ANDREASSEN2} A.~Andreassen, D.~Farhi, W.~Frost, M.D.~Schwartz,
{\sf arXiv:1604.06090}.

\bibitem{NOTE2} Essentially, these are also the conditions under 
which the resonant amplitude of a Gamow state can be identified with the
Breit-Wigner amplitude~\cite{NPA08}.


\bibitem{BESIII} BESIII Collaboration, {\sf arXiv:1612.05721}.

\bibitem{DETTORI} F.~Dettori (for the LHCb Collaboration),  
{\sf arXiv:1611.06717}.

\bibitem{LHCb1} LHCb Collaboration, JHEP~{\bf 03} (2016) 040
{\sf arXiv:1601.05284}.


\bibitem{LHCb2} LHCb Collaboration, Phys.~Lett.~B~{\bf 757}, 558 
(2016); {\sf arXiv:1510.08367}.

\bibitem{LHCb3} LHCb Collaboration, Eur.~Phys.~J.~C,~{\bf 77}, 72 (2017);
{\sf arXiv:1610.01383}.

\bibitem{LHCb4} LHCb Collaboration, Phys.~Rev.~D~{\bf 95}, 012006 (2017);
{\sf arXiv:1610.05187}.

\bibitem{BESIII2} BESIII Collaboration, 
Phys.~Rev.~Lett.~{\bf 117}, 042002 (2016); {\sf arXiv:1603.09653}.

\bibitem{ROTTER1} A.I.~Magunov, I.~Rotter, and S.I.~Strakhova, 
Phys.~Rev.~B~{\bf 68}, 245305 (2003); {\sf arXiv:quant-ph/0305064}.

\bibitem{ROTTER2} I.~Rotter, {\sf arXiv:0711.2926}.

\bibitem{LUNAACOSTA} G.A.~Luna-Acosta, A.A.~Fern\'andez-Mar{\'i}n, 
J.~A.~M\'endez-Berm\'udez, C.~Poli, Phys.~Lett.~A~{\bf 380}, 2494 (2016);
{\sf arXiv:1606.00326}.

\bibitem{BWM} For a q-extension of momentum Breit-Wigner distributions,
see Refs.~\cite{PLASTINO1,PLASTINO2}.

\bibitem{KAMANO15} H.~Kamano, S.X.~Nakamura, T.-S.H.~Lee, T.~Sato,
Phys.~Rev.~C~{\bf 92}, 025205 (2015); {\sf arXiv:1506.01768}.

\bibitem{MIYAHARA16} K.~Miyahara, T.~Hyodo, 
Phys.~Rev.~C~{\bf 93}, 015201 (2016); {\sf arXiv:1506.05724}.

\bibitem{KAMANOPRC16} H.~Kamano, S.X.~Nakamura, T.-S.H.~Lee, T.~Sato, 
Phys.~Rev.~C~{\bf 94}, 015201 (2016); {\sf arXiv:1605.00363}.


\bibitem{KAMANOPRC16b} H.~Kamano, T.-S.H.~Lee, 
Phys.~Rev.~C~{\bf 94}, 065205 (2016); {\sf arXiv:1608.03470}.

\bibitem{FLORES} A.~Flores-Tlalpa, G.~L\'opez Castro, P.~Roig,
J.~High Energy Phys.~{\bf 04}, 185 (2016); {\sf arXiv:1508.01822}.


       



\end{thebibliography}
\end{document}